\documentclass{article}
\usepackage[margin=3cm]{geometry}

\usepackage{amsmath,amssymb,amsfonts}
\usepackage{mathrsfs} %%% Needed for the manifold symbol M

\usepackage{tikz}
\usetikzlibrary{matrix}
\usepackage{setspace}

\usepackage{graphicx,color}
\usepackage{accents}

\usepackage[nocompress]{cite}

\newcommand{\ourG}{\mathbf{G}}
\newcommand{\ourB}{\mathbf{B}}
\newcommand{\ourC}{\mathbf{C}}

\numberwithin{equation}{section}

\usepackage{hyperref}
\hypersetup{colorlinks=true,breaklinks=true,linkcolor=blue,filecolor=magenta,urlcolor=cyan}
\title{Modified gravity: a unified approach}
\author{Christian G. B\"ohmer\footnote{Email: c.boehmer@ucl.ac.uk} \ and
  Erik Jensko\footnote{Email: erik.jensko.19@ucl.ac.uk} \\
  Department of Mathematics, University College London, \\
  Gower Street, London WC1E 6BT, United Kingdom}

\date{\today} 

\pagecolor{white}

\begin{document}

\maketitle

\begin{abstract}
  Starting from the original Einstein action, sometimes called the Gamma squared action, we propose a new setup to formulate modified theories of gravity. This can yield a theory with second order field equations similar to those found in other popular modified gravity models. Using a more general setting the theory gives fourth order equations. This model is based on the metric alone and does not require more general geometries. It is possible to show that our new theory and the recently proposed $f(Q)$ gravity models are equivalent at the level of the action and at the level of the field equations, provided that appropriate boundary terms are taken into account. Our theory can also reproduce $f(R)$ gravity, which is an expected result. Perhaps more surprisingly, we show that this equivalence extends to $f(T)$ gravity at the level of the action and its field equations, provided that appropriate boundary terms are taken in account. While these three theories are conceptually different and are based on different geometrical settings, we can establish the necessary conditions under which their field equations are the same. The final part requires matter to couple minimally to gravity. Through this work we emphasise the importance played by boundary terms which are at the heart of our approach.
\end{abstract}

\clearpage

\tableofcontents

\clearpage

\section{Introduction}

The field of modified or extended theories of gravity has become a substantial area of research since the early 2000s. Ever since, a plethora of new theories has been proposed and studied. However, the history of modifications of the original theory of General Relativity (GR) can be traced back to the 1920s. Shortly after GR was formulated it became apparent that the geometrical framework underlying the theory can be extended. Moreover, there was a parallel development on unifying the different fundamental forces into a single unified field theory, see~\cite{Goenner:2004se,Goenner:2014mka}.

These early geometrical attempts can all be studied under the umbrella of metric-affine theories of gravity~\cite{Hehl:1994ue}. Theories of this type typically are invariant under local Lorentz transformations and are also invariant under arbitrary coordinate transformations. When the Levi-Civita connection is replaced with the general affine connection, the matter action should also depend on this affine connection, thereby giving rise to possible source terms for torsion and non-metricity, in addition to the usual matter source terms for curvature like in GR\@. In case of the well-understood Einstein-Cartan theory~\cite{FH1976} one finds that mass couples to curvature while spin couples to torsion, see also~\cite{Hehl:1999bt}. Note that matter couplings to non-metricity can be more problematic~\cite{Neeman:1996zcr}. The field equations are second order in the metric but are algebraic in torsion which is a result of the specific action underlying the theory. By insisting on second order field equations for the outset, the specific form of the field equations is highly constrained~\cite{Lovelock:1971yv}.

The additional geometrical structure of metric-affine theories allows for other, equivalent, formulations of GR, the best known being the Teleparallel Equivalent of General Relativity (TEGR)~\cite{Nester1991,RA2013,Maluf:2013gaa}. For the present purpose, the most interesting fact about TEGR is that the action is not invariant under local Lorentz transformations but is pseudo-invariant. By this we mean that the action is invariant up to a boundary term. As this boundary term does not contribute to the resulting field equations these are then invariant. This is also true for the so-called Einstein action (or Gamma squared action) which differs from the usual Einstein-Hilbert action by a boundary term. Actions containing pseudo-invariant quantities will no longer yield invariant field equations if nonlinear functions of such quantities are considered, which brings us to another type of modified theories.

One of the earliest works on $f(R)$ gravity is probably~\cite{Barrow:1983rx}, where the stability of cosmological solutions was studied. It was, however, much later when these theories became popular, see~\cite{Sotiriou:2008rp,DeFelice:2010aj,Nojiri:2010wj}. These theories are invariant under local Lorentz transformations and diffeomorphisms, however, they lead to fourth order field equations in the metric. The Ricci scalar contains second derivatives of the metric so one would expect fourth order field equations already in GR; however, these higher order terms enter the action as a boundary term and so do not contribute to the field equations. If one were to separate the Ricci scalar into two parts, one of which only contains first derivatives of the metric, one would find a theory with second order equations, however, such a theory is likely to have other shortfalls. Many modified gravity models can be tightly constrained by cosmological observations~\cite{Clifton:2011jh,Koyama:2015vza} with GR generally emerging as the theory to provide the best fit to the data.

This is precisely the situation one arrives at when considering modifications of TEGR~\cite{Ferraro:2006jd,Bengochea:2008gz,Linder:2010py,Capozziello:2011et,Cai:2015emx,Nojiri:2017ncd,Heisenberg:2018vsk}. Here we are referring to the conventional formulation where the spin-connection is set to zero, but note that an alternative approach does exist, see for instance~\cite{Krssak:2018ywd}.

The action of TEGR only contains first derivatives of the tetrads and hence yields second order field equations. Consequently $f(T)$ also has second order field equations, however, one now loses invariance under local Lorentz transformations~\cite{Li:2010cg}. The modified action is no longer pseudo-invariant as it does not differ from an invariant action by a boundary term. Therefore, the field equations are no longer invariant under arbitrary Lorentz transformations, however, there exists nonetheless a remnant symmetry~\cite{Chen:2014qtl,Ferraro:2014owa} which affects the number of degrees of freedom of $f(T)$ gravity~\cite{Li:2011rn,Ferraro:2018tpu,Ferraro:2018axk,Ferraro:2020tqk,Blagojevic:2020dyq}.

The important role played by boundary terms was investigated in~\cite{Bahamonde:2015zma,Bahamonde:2016kba,Bahamonde:2017wwk} where an arbitrary function depending on the torsion scalar $T$ and the relevant boundary term $B$ was considered, $f(T,B)$ gravity. Within this framework it is possible to identify $f(R)$ gravity as the unique Lorentz invariant theory and to identify $f(T)$ gravity as the unique second order theory. 

Similarly to TEGR one can construct another alternative formulation of GR which is based on non-metricity, instead of torsion~\cite{Nester:1998mp,Poltorak:2004tz,Adak:2005cd,Vitagliano:2013rna,Mol:2014ooa,Jarv:2018bgs}. Some constraints on spacetime non-metricity can be found in~\cite{Foster:2016uui,Lehnert:2017tep}. If we denote the relevant non-metricity scalar by $Q$ it is natural to consider generalised theories which are called $f(Q)$ gravity~\cite{BeltranJimenez:2017tkd,Harko:2018gxr,Soudi:2018dhv,BeltranJimenez:2019tjy,Jimenez:2019ovq}. Theories of this type are, by construction, invariant under Lorentz transformations. Generally one works in the so called coincidence gauge~\cite{Hohmann:2018wxu,Dialektopoulos:2019mtr,Harko:2018gxr,Jimenez:2019ovq,Xu:2020yeg} where it is important to study the conservation equation carefully. 

In GR the energy-momentum conservation equation is implied by Noether's theorem and follows from the invariance of the Einstein-Hilbert action under arbitrary coordinate transformations~\cite{DN:2017}, see also~\cite{Bak:1993us,Wald:1999wa,Padmanabhan:2004xk,Schrodinger:2011gqa,1010070-8176-4454-78,Obukhov:2014nja,Obukhov:2015eha,DeHaro:2021gdv}. From a geometrical point of view one generally argues that the conservation equation is a consequence of the twice contracted Bianchi identities, however, in modified theories of gravity it is desirable to follow Noether's approach. In certain situations the Bianchi identity can appear in a somewhat unusual form, as in $f(Q)$ gravity mentioned above, but also in teleparallel gravity~\cite{MOLLER1958347,Moller:1962vis,Chen:2015vya}. Also of interest are the recently proposed minimally modified theories~\cite{Lin:2017oow,Carballo-Rubio:2018czn} in which the theory is invariant under spatial diffeomorphisms only.

When matter is coupled non-minimally to the gravitational field~\cite{Bertolami:2007gv,Harko:2018ayt,Carloni:2016glo,Boehmer:2018zrm} test particles do not necessarily follow geodesic motion and one can also construct models where the energy-momentum tensor is no longer conserved~\cite{Charmousis:2009tc,Capozziello:2014bqa,Boehmer:2013ss,Boehmer:2014ipa}. Such models often have close links to approaches where the theory is not fully invariant under all coordinate transformations~\cite{Anber:2009qp,Bluhm:2014oua,Bluhm:2016dzm,Velten:2021xxw}. Theories which break local Lorentz or diffeomorphism invariance can be motivated by quantum gravity considerations. If we accept that classical field theories like GR cannot be applied at length scales where quantum effects dominate, then it becomes natural to consider models that break certain symmetries at small scales. The so-called Born-Infeld scheme~\cite{Born:1934gh} is one of the simplest approaches to achieve this and has been successfully implemented in $f(T)$ gravity~\cite{Ferraro:2006jd,Ferraro:2008ey,Fiorini:2009ux,Boehmer:2019uxv,Boehmer:2020hkn}. A completely different approach would be to reformulate differential geometry to take into account an underlying quantum structure~\cite{Majid:2020}.

Our model, which will be based on an arbitrary function of two coordinate pseudo scalars, contains all of the above issues to certain degrees and we will carefully address them. While our theory is based on coordinate pseudo scalars, these are scalars which are not invariant under diffeomorphisms, we will nonetheless arrive at field equations which are tensor-like. After discussing infinitesimal coordinate transformations of our action, we are able to formulate a consistent theory which reduces to GR in the appropriate limit. When studying our model in a cosmological setting, we find the same field equations that were derived in $f(T)$ gravity and in $f(Q)$ gravity. This observation leads the way towards studying the equivalence of these theories. In particular, when comparing the relevant field equations, rewritten conveniently, one is tempted to declare their equivalence at once. However, this would be premature as various boundary terms need to be taken into account in order to discuss these theories in a unified setting. The main result of this work is to identify three boundary terms with unusual properties that allows us to construct one general family of modified gravity theories which will contain the various aforementioned models as limiting cases.

Our paper is organised as follows: in Section~\ref{sec:EH} we discuss the Einstein-Hilbert and the Gamma squared action together with coordinate transformations. Section~\ref{sec:mod} introduces our model, discusses its various properties and establishes equivalence with $f(R)$ in a certain limit. The following Section~\ref{sec:equiv} contains the required details to show the equivalence of our model with other modified theories of gravity, a summary of this part is given in Fig.~\ref{fig2}. We move onto a discussion of our work in Section~\ref{sec:disc}. A comprehensive appendix is provided containing the variations of the actions.

\subsection*{Notation}

Throughout this paper the signature is $(-,+,+,+)$, Greek indices are spacetime indices taking values $(0,1,2,3)$ and Latin indices denote tangent space indices. Our co-frame and frame fields or tetrads are defined via $g_{\mu\nu} = e_\mu^a e_\nu^b \eta_{ab}$ and $g^{\mu\nu} = E^\mu_a E^\nu_b \eta^{ab}$ with $E^\mu_a e_\mu^b = \delta_a^b$ and $E^\mu_a e_\nu^a = \delta_\nu^\mu$. $g$ is the determinant of the metric $g=\det g_{\mu\nu}$ and we denote $e=\det e_\mu^a$ so that $e=\sqrt{|g|}$. Round brackets denote symmetrisation and square brackets denote skew-symmetrisation: $t_{(\mu\nu)} = 1/2(t_{\mu\nu}+t_{\nu\mu})$ and $t_{[\mu\nu]} = 1/2(t_{\mu\nu}-t_{\nu\mu})$. We set $G = c = 1$ and use $\kappa=8\pi$ as the gravitational coupling constant.

\section{The Einstein-Hilbert and Gamma squared actions}
\label{sec:EH}

\subsection{The actions}

Let us begin with the standard Einstein-Hilbert action of General Relativity,
\begin{align}
  \label{EH_Action}
  S_{\rm EH}[g_{\mu \nu}] = \frac{1}{2 \kappa} \int R \sqrt{-g}\, d^4x \,,
\end{align}
where $R$ stands for the Ricci scalar or curvature scalar. The square brackets indicate explicitly the dynamical variables of the theory. It has been know since the formulation of GR that the Ricci scalar $R$ can be rewritten as a `bulk' term $\ourG$ and a boundary or surface term $\ourB$. Using this decomposition, the Einstein-Hilbert action becomes
\begin{align}
  \label{EH_G_B_Action}
  S_{\rm EH}[g_{\mu \nu}] =\frac{1}{2 \kappa}  \int R \sqrt{-g} \, d^4x =
  \frac{1}{2 \kappa}  \int \bigl( \ourG + \ourB \bigr) \sqrt{-g} \, d^4x \,.
\end{align}
The bulk term $\ourG$ is quadratic in the connection coefficients or Christoffel symbols components and first order in the metric derivatives
\begin{align}
  \label{G_bulk}
  \ourG &= g^{\mu \nu}\Big( \Gamma^{\lambda}_{\mu \sigma} \Gamma^{\sigma}_{\lambda \nu} -
  \Gamma^{\sigma}_{\mu \nu} \Gamma^{\lambda}_{\lambda \sigma} \Big) \,.
\end{align}
We note for future reference that term in the bracket is not symmetric by construction, however, it is clear that only its symmetric part contributes to the action due to $g^{\mu\nu}$. As the Christoffel symbols appear in the geodesic equation, they can be seen as the gravitational force terms which determine the motion of test particles. This is similar to the Faraday tensor in the Lorentz force equation. It is thus natural to consider a Lagrangian quadratic in these quantities which makes $\ourG$ an appealing choice. The boundary term $\ourB$ is second order in the metric derivatives
\begin{align}
  \label{B_boundary}
  \ourB =  \frac{1}{\sqrt{-g}} \partial_{\nu}
  \Bigl( \frac{\partial_{\mu}(g g^{\mu \nu})}{\sqrt{-g}} \Bigr) =
  \frac{1}{\sqrt{-g}} \partial_{\sigma} (\sqrt{-g} B^{\sigma}) = \nabla_{\sigma} B^{\sigma} \,,
\end{align}
where we introduce the notation 
\begin{align}
  \label{B_boundary_vector}
  B^{\sigma} = g^{\mu \nu} \Gamma^{\sigma}_{\mu \nu} - g^{\sigma \nu} \Gamma^{\lambda}_{\lambda \nu} \,.
\end{align}
Before continuing let us make some remarks: at this point, we are working in curved spacetime with vanishing torsion and vanishing non-metricity so that the general affine connection is the usual Levi-Civita one. Note that neither $\ourG$ nor $\ourB$ transform like scalars under general coordinate transformations (diffeomorphisms) as they are not true scalars. We will refer to such scalars as \emph{pseudo scalars}. It is only when combining these two objects into the Ricci scalar that one retrieves a true scalar, an object invariant under general coordinate transformations. For this reason, one has to be careful when working with these quantities. Nonetheless, it is convenient to refer to $B^{\sigma}$ as the boundary vector despite is non-vectorial nature. Since the term $\ourB$ takes the form of a total derivative, it will not appear in the Euler-Lagrange equations provided we are working on a manifold without boundary. In the presence of boundaries we would assume variations to vanish on those boundaries.

Consequently, the Einstein field equations derived from the Einstein-Hilbert action~(\ref{EH_G_B_Action}) arise solely from the $\ourG$ term containing only first derivatives. It is common to refer to this as the \emph{Einstein action} or the \emph{Gamma squared action}
\begin{align}
  \label{Einstein_Action}
  S_{\rm E} [g_{\mu \nu}] =\frac{1}{2 \kappa}  \int \ourG \sqrt{-g} \, d^4x \,.
\end{align}
As already touched upon, this action is manifestly non-covariant since $\ourG$ is not a true scalar under general coordinate transformations. However, the action is diffeomorphism invariant up to a boundary term, in which case one often speaks of pseudo-invariance. We will look at this explicitly shortly, as the results will be useful in the following sections. 

For completeness Appendix~\ref{Appendix_Einstein} contains the derivation of the Einstein field equations when working directly with $S_{\rm E}[g_{\mu \nu}]$, along with the minimally coupled matter action describing the matter fields $S_{\rm matter}[g_{\mu \nu},\Phi]$. Some of the results contained in that appendix will be required for subsequent calculations. Let us introduce the following two non-tensorial objects
\begin{align}
  \label{M}
  M^{\mu \nu}{}_{\lambda} &:=
  \frac{\delta \ourG}{\delta \Gamma^{\lambda}_{\mu \nu}} =
  2 g^{\rho (\nu} \Gamma^{\mu)}_{\lambda \rho} - g^{\mu \nu} \Gamma^{\rho}_{\rho \lambda} -
  g^{\rho \sigma}  \delta^{(\nu}_{\lambda} \Gamma^{\mu)}_{\rho \sigma} \, , \\
  \label{E}
  E^{\mu \nu \lambda}  &:= M^{\{ \lambda \mu \nu \}} =
  M^{\lambda \mu \nu} + M^{\nu \lambda \mu} - M^{\mu \nu \lambda} \nonumber \\
  &= 2 g^{\rho \mu} g^{\nu \sigma} \Gamma^{\lambda}_{\rho \sigma} -
  2 g^{\lambda (\mu} g^{\nu) \sigma} \Gamma^{\rho}_{\rho \sigma} + g^{\mu \nu} g^{\lambda \rho} 
  \Gamma^{\sigma}_{\sigma \rho} - g^{\mu \nu} g^{\rho \sigma} \Gamma^{\lambda}_{\rho \sigma} \,.
\end{align}
The first object naturally follows from the variation of the bulk term $\ourG$ with respect to the connection. The second one has the same linear index combination that appears in the definition of the Christoffel symbol. The notation $\{\ldots\}$ with three indices is sometimes called the Schouten bracket, see~\cite{JS1954}. Both terms~(\ref{M}) and~(\ref{E}) are symmetric over their first two indices, which follows directly from the symmetry of the connection. Note that using definition~(\ref{M}) we can write the bulk term as
\begin{equation}
  \ourG = \frac{1}{2} M^{\mu \nu}{}_{\lambda} \Gamma^{\lambda}_{\mu \nu} \,.
\end{equation}
The boundary vector defined in equation~(\ref{B_boundary_vector}) can be obtained by the contraction $E^{\mu}{}_{\mu}{}^{\sigma} = -2B^{\sigma}$.

\subsection{Coordinate transformations}
\label{Coord_transformations}

Let us determine how the newly introduced objects transform under infinitesimal coordinate transformations $x^{\mu} \rightarrow \hat{x}^{\mu} = x^{\mu} + \xi^{\mu}(x)$, where $\xi$ is assumed to be small, $|\xi^{\mu}| \ll 1$. To first order in $\xi^{\mu}$, we have the following relations
\begin{align}
  \frac{\partial \hat{x}^{\mu}}{\partial x^{\nu}} = \delta^{\mu}_{\nu} + \partial_{\nu} \xi^{\mu} \,,
  \qquad
  \frac{\partial x^{\mu}}{\partial \hat{x}^{\nu}} = \delta^{\mu}_{\nu} - \partial_{\nu} \xi^{\mu} \,.
\end{align}
Using the above, along with the usual transformation laws, the metric and Christoffel symbol transform as
\begin{align}
  \hat{g}_{\mu \nu}(\hat{x}) &= g_{\mu \nu} - \partial_{\mu} \xi^{\lambda} g_{\lambda \nu} -
  \partial_{\nu} \xi^{\lambda} g_{\mu \lambda} + \mathcal{O} (\xi^2) \,, \\
  \hat{g}^{\mu \nu}(\hat{x})  &= g^{\mu \nu} + \partial_{\lambda} \xi^{\mu} g^{\lambda \nu} +
  \partial_{\lambda} \xi^{\nu} g^{\mu \lambda} + \mathcal{O} (\xi^2) \,, \\
  \hat{\Gamma}^{\gamma}_{\mu \nu}(\hat{x})  &= \Gamma^{\gamma}_{\mu \nu} + \partial_{\lambda} \xi^{\gamma} \Gamma^{\lambda}_{\mu \nu} - \partial_{\mu} \xi^{\lambda} \Gamma^{\gamma}_{\nu \lambda} - \partial_{\nu} \xi^{\lambda} \Gamma^{\gamma}_{\mu \lambda} -\partial_{\mu} \partial_{\nu} \xi^{\gamma} +  \mathcal{O} (\xi^2)  \,.
\end{align}
The terms on the right-hand sides are functions of the original coordinates $x^{\mu}$.

Using these transformations in our definition of $\ourG$~(\ref{G_bulk}), and dropping the higher order $\xi$ terms, one finds through simple computation
\begin{align}
  \label{G_infinitesimal}
  \hat{\ourG}(\hat{x})  &=
  \hat{g}^{\mu \nu}\big( \hat{\Gamma}^{\lambda}_{\mu \sigma} \hat{\Gamma}^{\sigma}_{\lambda \nu} -
  \hat{\Gamma}^{\sigma}_{\mu \nu} \hat{\Gamma}^{\lambda}_{\lambda \sigma} \big) \nonumber \\
  &= g^{\mu \nu}\big( \Gamma^{\lambda}_{\mu \sigma} \Gamma^{\sigma}_{\lambda \nu} -
  \Gamma^{\sigma}_{\mu \nu} \Gamma^{\lambda}_{\lambda \sigma} \big) - \big(2 g^{\mu (\alpha} \Gamma^{\beta)}_{\mu \gamma} - g^{\alpha \beta} \Gamma^{\lambda}_{\lambda \gamma} - g^{\mu \nu} \delta^{(\beta}_{\gamma} \Gamma^{\alpha)}_{\mu \nu} \big) \partial_{\alpha} \partial_{\beta} \xi^{\gamma}  \nonumber \\
  &= \ourG - M^{\alpha \beta}{}_{\gamma} \partial_{\alpha} \partial_{\beta} \xi^{\gamma} \,,
  \end{align}
with $M^{\alpha \beta}{}_{\gamma}$ defined in~(\ref{M}). A similar calculation for the boundary vector~(\ref{B_boundary_vector}) leads to
\begin{align}
\hat{B}^{\sigma}(\hat{x})  =  B^{\sigma} + \partial_{\eta} \xi^{\sigma} B^{\eta} - \partial^{\nu} \partial_{\nu} \xi^{\sigma} + \partial^{\sigma} \partial_{\lambda} \xi^{\lambda} \,,
\end{align}
and for the full boundary term~(\ref{B_boundary}) we find
\begin{align}
  \label{B_infinitesimal} 
  \hat{\ourB}(\hat{x})  &=
  \hat{\Gamma}^{\lambda}_{\lambda \sigma} \hat{B}^{\sigma} + \hat{\partial}_{\sigma} \hat{B}^{\sigma}
  \nonumber \\
  &= \ourB +  \big(2 g^{\mu (\alpha} \Gamma^{\beta)}_{\mu \gamma} -
  g^{\alpha \beta} \Gamma^{\lambda}_{\lambda \gamma} - g^{\mu \nu}
  \delta^{(\beta}_{\gamma} \Gamma^{\alpha)}_{\mu \nu} \big) \partial_{\alpha} \partial_{\beta} \xi^{\gamma}
  \nonumber \\
  &=  \ourB + M^{\alpha \beta}_{}{\gamma} \partial_{\alpha} \partial_{\beta} \xi^{\gamma} \,.
\end{align}
Equations~(\ref{G_infinitesimal}) and~(\ref{B_infinitesimal}) of course imply that $\hat{\ourG} + \hat{\ourB} = \ourG + \ourB$ as expected. This follows from the fact that $R=\ourG + \ourB$ is the Ricci scalar.

For completeness we also compute the infinitesimal coordinate transformations for the three-index objects $M^{\alpha\beta}{}_{\gamma}$ and $E_{\rho \sigma}{}^{\gamma}$. For the former we find
\begin{multline}
  \label{M_infinitesimal}
  \hat{M}^{\alpha \beta }{}_{\gamma}  =
  M^{\alpha \beta }{}_{\gamma} + \partial_{\lambda} \xi^{\alpha } M^{\lambda \beta }{}_{\gamma} +
  \partial_{\lambda} \xi^{\beta } M^{\alpha  \lambda}{}_{\gamma} - \partial_{\gamma} \xi^{\lambda} M^{\alpha  \beta }{}_{\lambda} \\
  - 2 \partial^{(\alpha }\partial_{\gamma} \xi^{\beta )} + g^{\alpha \beta } \partial_{\lambda} \partial_{\gamma} \xi^{\lambda} +
  g^{\mu \nu} \delta^{(\beta }_{\gamma} \partial_{\mu} \partial_{\nu} \xi^{\alpha )} \,,
\end{multline}
while the latter gives
\begin{multline}
  \label{E_infinitesimal}
  \hat{E}_{\rho \sigma}{}^{\gamma}  =  E_{\rho \sigma}{}^{\gamma} -
  E_{\rho \eta}{}^{\gamma} \partial_{\sigma} \xi^{\eta} -
  E_{\sigma \eta}{}^{\gamma} \partial_{\rho} \xi^{\eta} +
  E_{\rho \sigma}{}^{\eta} \partial_{\eta} \xi^{\gamma} \\ 
  - 2 \partial_{\rho} \partial_{\sigma} \xi^{\gamma} +
  2 \delta_{(\rho}^{\gamma} \partial_{\sigma)} \partial_{\lambda} \xi^{\lambda} -
  g_{\rho \sigma} \partial^{\gamma} \partial_{\lambda} \xi^{\lambda} +
  g_{\rho \sigma} \partial^{\lambda} \partial_{\lambda} \xi^{\gamma} \,.
\end{multline}

Using the calculations above, we can state the Lie derivatives\footnote{Specifically, $\mathcal{L}_{\xi} T(x) =  \lim_{\epsilon \rightarrow 0}  \frac{1}{\epsilon}\big(T(x)-\hat{T}(x) \big)$, where $\epsilon$ has previously been absorbed into the definition of $\xi^{\mu}$ and $\hat{T}(x)$ represents the transformed object pulled back to the original coordinates $x^{\mu}$. For brevity we omit all factors of $\epsilon$ and simply write $\mathcal{L}_{\xi} T(x) = T(x)-\hat{T}(x)$, which we understand to mean the former definition.}. For the metric and the connection we get the well-known \cite{JS1954,Wald:1984,Yano:2020} results
\begin{align}
  \mathcal{L}_{\xi} g_{\mu \nu} = g_{\mu \nu} - \hat{g}_{\mu \nu} =
  \partial_{\lambda} {g}_{\mu \nu} \xi^{\lambda} + \partial_{\mu} \xi^{\lambda} g_{\lambda \nu} +
  \partial_{\nu} \xi^{\lambda} g_{\mu \lambda}  = 2 \nabla_{(\mu} \xi_{\nu)} \,,
\end{align}
and
\begin{align}
  \label{Lie_Gamma}
  \mathcal{L}_{\xi} \Gamma^{\gamma}_{\mu \nu} &=
  \Gamma^{\gamma}_{\mu \nu} - \hat{\Gamma}^{\gamma}_{\mu \nu} =
  \partial_{\lambda} \Gamma^{\gamma}_{\mu \nu} \xi^{\lambda} - \partial_{\lambda} \xi^{\gamma} \Gamma^{\lambda}_{\mu \nu} +\partial_{\mu} \xi^{\lambda} \Gamma^{\gamma}_{\nu \lambda} + \partial_{\nu} \xi^{\lambda} \Gamma^{\gamma}_{\mu \lambda} + \partial_{\mu} \partial_{\nu} \xi^{\gamma} \nonumber \\
&= \nabla_{\mu} \nabla_{\nu} \xi^{\gamma} + R_{\rho \mu \nu}{}^{\gamma} \xi^{\rho} \,.
\end{align}

Interestingly the Lie derivative of the connection is a tensorial quantity which follows from the fact that the difference of two connections is always a tensor. This will not be true for our objects $\ourG$ and $\ourB$ though because we have products of Gammas that cannot be rewritten in terms of covariant derivatives or other tensors. Consequently, for the bulk term $\ourG$ we have
\begin{align}
  \label{Lie_G}
  \mathcal{L}_{\xi} \ourG = \xi^{\mu}  \partial_{\mu} \ourG +
  M^{\alpha \beta}{}_{\gamma} \partial_{\alpha} \partial_{\beta} \xi^{\gamma} \,,
\end{align}
and for the boundary term $\ourB$ 
\begin{align} \label{Lie_B}
\mathcal{L}_{\xi} \ourB = \xi^{\mu} \partial_{\mu} \ourB - M^{\alpha \beta}{}_{\gamma} \partial_{\alpha} \partial_{\beta} \xi^{\gamma} \,.
\end{align}
Let us note that these relation can be derived in two equivalent ways. First, one can derive the transformation property under infinitesimal coordinate transformations and compute their difference. Second, one could use the known expressions for the Lie derivative and apply those directly.

Finally, we also state the relevant expression for $M$ which reads
\begin{multline} 
  \label{Lie_M}
  \mathcal{L}_{\xi} M^{\alpha \beta }{}_{\gamma} = \xi^{\mu} \partial_{\mu} M^{\alpha \beta }{}_\gamma -
  2 \partial_{\lambda} \xi^{(\alpha } M^{\beta ) \lambda}{}_{\gamma}  + \partial_{\gamma} \xi^{\lambda} M^{\alpha \beta }{}_{\lambda} \\
  + 2 \partial^{(\alpha }\partial_{\gamma} \xi^{\beta )} - g^{\alpha \beta } \partial_{\lambda} \partial_{\gamma} \xi^{\lambda} -
  g^{\mu \nu} \delta^{(\beta }_{\gamma} \partial_{\mu} \partial_{\nu} \xi^{\alpha )} \,.
\end{multline}
Similarly for $E$ one arrives at
\begin{multline}
  \label{Lie_E}
  \mathcal{L}_{\xi} {E}_{\rho \sigma}{}^{\gamma} =  \xi^{\mu} \partial_{\mu} E_{\rho \sigma}{}^{\gamma} +
  2 E_{\eta (\rho}{}^{\gamma} \partial_{\sigma)} \xi^{\eta} -
  E_{\rho \sigma}{}^{\eta} \partial_{\eta} \xi^{\gamma} \\ 
  + 2 \partial_{\rho} \partial_{\sigma} \xi^{\gamma} - 2 \delta_{(\rho}^{\gamma} \partial_{\sigma)}
  \partial_{\lambda} \xi^{\lambda} + g_{\rho \sigma} \partial^{\gamma} \partial_{\lambda} \xi^{\lambda} -
  g_{\rho \sigma} \partial^{\lambda} \partial_{\lambda} \xi^{\gamma} \,.
\end{multline}
The right-hand sides of equations~(\ref{Lie_G})--(\ref{Lie_E}) are all non-tensorial. However, this is not immediately obvious as one might also not immediately identify $\mathcal{L}_{\xi} \Gamma$ to be tensorial. The easiest way to see is to recall that $\ourG$ is quadratic in the Christoffel symbol components, so that symbolically $\mathcal{L}_{\xi} \Gamma^2 \sim \Gamma \mathcal{L}_{\xi} \Gamma$ which cannot be a tensor as it is a product of a tensor and a connection.

\subsection{Diffeomorphism invariance}
\label{Einstein_conservation}
  
As shown in Appendix~\ref{Appendix_Einstein} the variation of the Einstein action with respect to the metric leads to
\begin{align} 
  \delta S_{E}[g_{\mu \nu}] = \frac{1}{2 \kappa}  \int \delta{g^{\mu \nu}} G_{\mu \nu} \sqrt{-g} d^4 x \,.
  \label{S_var}
\end{align}
Let us now consider the variation of the action under a diffeomorphism generated by the infinitesimal vector field $\xi$, denoted by $\delta_\xi$. The most direct approach is to make use of the Lie derivatives introduced in the previous subsection, thus we find
\begin{align}
  \delta_{\xi} S_{E}[g_{\mu \nu}] &=
  \frac{1}{2 \kappa}\int \mathcal{L}_{\xi} \big(  \sqrt{-g} \ourG ) d^4 x  =
  \frac{1}{2 \kappa}\int \mathcal{L}_{\xi} (\sqrt{-g}) \ourG +
  \sqrt{-g} \mathcal{L}_{\xi}(\ourG)  \ d^4 x  \nonumber \\
  &= \frac{1}{2 \kappa} \int \frac{1}{2} \sqrt{-g} g^{\mu \nu} (\mathcal{L}_{\xi} g_{\mu \nu}) \ourG +
  \sqrt{-g} \mathcal{L}_{\xi}(\ourG)  \ d^4 x  \nonumber \\
  &= \frac{1}{2 \kappa}\int \sqrt{-g} (\nabla_{\mu} \xi^{\mu}) \ourG +
  \sqrt{-g} \Big(\partial_{\mu} (\ourG) \xi^{\mu} + M^{\alpha \beta}{}_{\gamma}
  \partial_{\alpha} \partial_{\beta} \xi^{\gamma}\Big) \ d^4 x \nonumber \\
  &= \frac{1}{2 \kappa}\int \sqrt{-g} \partial_{\mu} \xi^{\mu} \ourG +
  \sqrt{-g} \Gamma^{\mu}_{\mu \nu} \xi^{\nu} \ourG + \sqrt{-g} \partial_{\mu} \ourG \xi^{\mu} +
  \sqrt{-g} M^{\alpha \beta}{}_{\gamma} \partial_{\alpha} \partial_{\beta} \xi^{\gamma}\, d^4 x
  \nonumber \\
  &= \frac{1}{2 \kappa}\int \partial_{\mu} \big( \sqrt{-g} \xi^{\mu} \ourG \big) +
  \sqrt{-g} M^{\alpha \beta}{}_{\gamma} \partial_{\alpha} \partial_{\beta} \xi^{\gamma}\, d^4 x \,.
  \label{eqn:f0}
\end{align}
Using integration by parts we can rewrite the final term as follows,
\begin{align}
  M^{\alpha \beta}{}_{\gamma} \partial_{\alpha} \partial_{\beta} \xi^{\gamma} = \textrm{boundary terms} +
  \frac{1}{\sqrt{-g}} \partial_{\alpha} \partial_{\beta}
  (\sqrt{-g} M^{\alpha\beta}{}_{\gamma}) \xi^{\gamma} \,,
  \label{eqn:f1}
\end{align}
where we use the generic expression `boundary terms' for all terms which appear that do not contribute to the equations of motion. Throughout this paper we are working on manifolds without boundaries. Should one wish to generalise these results to manifolds with boundaries, one could for instance assume that all variations vanish on that boundary. Alternatively, one could subtract suitable boundary terms from the action. We will not discuss the latter. 

It is interesting to note that the final term of~(\ref{eqn:f1}) is in fact proportional to the twice contracted Bianchi identities, albeit written in an unusual way that is not manifestly covariant~\cite{MOLLER1958347,Chen:2015vya,Duan:2018fke}. This result is perhaps expected as we are working with General Relativity after all. Hence, we find that upon integrating by parts twice
\begin{align}
  \delta_{\xi} S_{E}[g_{\mu \nu}] = \textrm{boundary terms} + \frac{1}{2 \kappa}
  \int  \partial_{\alpha} \partial_{\beta} (\sqrt{-g} M^{\alpha\beta}{}_{\gamma}) \xi^{\gamma}  d^4x = 0 \,,
\end{align}
where the last term vanishes identically. Therefore our action is diffeomorphism invariant, up to boundary terms. Unsurprisingly, the twice contracted Bianchi identity appears just like it does when considering the diffeomorphism invariance of the usual Einstein-Hilbert action. There is one subtle point to note though; here it is not seen as a consequence of diffeomorphism invariance but the reason for it.

The more elegant route leading to the same conclusion would have been to take the variation of the action with respect to the metric to obtain the Einstein tensor~(\ref{S_var}), then replace $\delta_{\xi} g^{\mu \nu}$ with the Lie derivative and again use the contracted Bianchi-identity to show that this vanishes. See for instance~\cite{TP:2010} where this is explicitly discussed. However, this more convoluted method, calculating the Lie derivatives explicitly, will be useful in the next section when considering generalisations of the action and the quicker route cannot be taken.

Before proceeding, recall that the variation of an arbitrary matter action is given by 
\begin{align} \label{matter_action_variation}
  \delta S_{\rm{matter}}[g_{\mu \nu}, \Phi] = \int \frac{\delta L_{\rm{M}}}{\delta g^{\mu \nu} }\delta g^{\mu \nu} \, d^4 x + \int \frac{\delta L_{\rm{M}}}{\delta \Phi} \delta \Phi \, d^4 x  \,,
\end{align}
where $\Phi$ denotes the matter fields and $L_{\rm{M}}$ is the matter Lagrangian (density). Let us assume $\Phi$ satisfies the matter equations of motion, which we will do throughout, such that the second term vanishes,
\begin{align}
\delta S_{\rm{matter}}[g_{\mu \nu}, \Phi] = \int \frac{\delta L_{\rm{M}}}{\delta g^{\mu \nu} }\delta g^{\mu \nu} \, d^4 x = \frac{1}{2} \int \sqrt{-g} T^{\mu \nu} \delta g_{\mu \nu} \, d^4x \,,
\end{align}
with the metric energy-momentum tensor defined as
\begin{align}
  T_{\mu \nu} = -\frac{2}{\sqrt{-g}}\frac{\delta L_{\rm{M}}[g_{\mu \nu}, \Phi]}{\delta g^{\mu \nu}} \,.
\end{align}
If we consider the variation resulting from an arbitrary diffeomorphism we have
\begin{align} \label{eqn:matter}
\delta_{\xi} S_{\rm{matter}}  &=  \frac{1}{2} \int \sqrt{-g} T^{\mu \nu} \mathcal{L}_{\xi} g_{\mu \nu}  \, d^4 x  = \int \sqrt{-g} T^{\mu \nu} \nabla_{\mu} \xi_{\nu}   \, d^4 x \nonumber \\
&= \textrm{boundary terms} - \int \sqrt{-g} \nabla_{\mu}(T^{\mu \nu}) \xi_{\nu} \, d^4 x \,.
\end{align}
One generally assumes that the matter action is invariant under coordinate transformations, which yields the conservation equation $\nabla_\mu T^{\mu\nu} = 0$. Likewise, assuming that the total action $S_{\rm total} = S_{\rm{E}} + S_{\rm matter}$ is invariant (not necessarily its individual parts) also yields the conservation equation $\nabla_\mu T^{\mu\nu} = 0$, as the twice contracted Bianchi identities ensure the vanishing of the geometrical part. This final statement is often to referred to as Noether's theorem of General Relativity~\cite{DN:2017}. We can write this as
\begin{align}
  \delta_{\xi} S_{\rm total} =
  \int \biggl[\frac{1}{2 \kappa} \partial_{\alpha} \partial_{\beta} (\sqrt{-g} M^{\alpha\beta}{}_{\gamma}) - \sqrt{-g} \nabla_\alpha T^{\alpha}_{\gamma} \biggr]\xi^\gamma \, d^4x \,,
  \label{eqn:f2}
\end{align}
and note again its slightly unusual form compared with the standard formulation.

\section{The modified Einstein action}
\label{sec:mod}

\subsection{\texorpdfstring{$f(\ourG, \ourB)$}{f(G,B)} gravity}

Generalising the results of the previous section, it is interesting to look at the class of theories where we consider arbitrary functions of $\ourG$ and $\ourB$, analogous to in $f(R)$ gravity. More closely related are theories of modified teleparallel gravity, where the Lagrangian $f(T,B)$ is a function of the torsion scalar and its boundary term \cite{Li:2010cg,Bahamonde:2015zma}. In this case $T$ and $B$ are not Lorentz-invariant, but the combination $-T+B$ gives the Ricci scalar, which of course is both a Lorentz scalar and a coordinate scalar. For our case with $f(\ourG,\ourB)$, as previously explained, the individual terms $\ourG$ and $\ourB$ are not diffeomorphism invariant except in the combination $\ourG + \ourB = R$. We will look more closely at the relation to other geometric formulations of gravity in Section~\ref{sec:equiv}.

The $f(\ourG, \ourB)$ class of theories encompass both General Relativity and $f(R)$ gravity for specific forms of the function $f$. However, they also give room for deviations by breaking the diffeomorphism symmetry present in those two cases. Consequently, let us consider the following gravitational $f(\ourG,\ourB)$ action
\begin{equation} \label{f(G,B)_action}
    S_{\rm grav}[g_{\mu \nu}] = \frac{1}{2 \kappa} \int f(\ourG,\ourB) \sqrt{-g} \ d^4 x \,. 
\end{equation}
Varying the action gives
\begin{equation} \label{f(G,B)_action_variation}
    \delta S_{\rm grav} = \frac{1}{2 \kappa} \int \left[ \delta  f(\ourG,\ourB) \sqrt{-g} - \frac{1}{2}  f(\ourG,\ourB)\sqrt{-g} g_{\rho \sigma} \delta g^{\rho \sigma} \right]  d^4 x \,,
\end{equation}
where
\begin{equation} \label{f(G,B)_variation}
    \delta  f(\ourG,\ourB) = \frac{\partial  f(\ourG,\ourB)}{\partial \ourG} \delta \ourG + \frac{\partial  f(\ourG,\ourB)}{\partial \ourB} \delta \ourB \,,
\end{equation}
and $\ourG$ and $\ourB$ are defined in~(\ref{G_bulk}) and~(\ref{B_boundary}). To calculate the variation we follow the same procedure as in Appendix~\ref{Appendix_Einstein}, using the objects previously introduced, $M^{\mu\nu\lambda}$~(\ref{M}) and $E^{\mu\nu\lambda}$~(\ref{E}). The full derivation is given in Appendix~\ref{Appendix_f(G,B)_action}. The resulting gravitational field equations of the total action are
\begin{multline}
  \label{f(g,b)_final_form}
     \frac{\partial f}{\partial \ourG} \Big[G_{\rho \sigma} + \frac{1}{2} g_{\rho \sigma} \ourG \Big] + \frac{1}{2}  E_{\rho \sigma}{}^{\gamma} \partial_{\gamma} \Big( \frac{\partial f}{\partial \ourG} \Big)  - \frac{1}{2} g_{\rho \sigma}  f(\ourG,\ourB) \ 
      \\
    + \frac{1}{2} \frac{\partial f}{\partial \ourB}  \ g_{\rho \sigma} \ourB  + g_{\rho \sigma}\partial^{\mu}  \partial_{\mu}\Big( \frac{\partial f}{\partial \ourB}\Big)
      - \partial_{\rho} \partial_{\sigma} \Big( \frac{\partial f}{\partial \ourB}\Big)
    \\
    + \frac{1}{2} g_{\rho \sigma} \partial_{\mu}(g^{\mu \nu}) \partial_{\nu}\Big(\frac{\partial f}{\partial \ourB}\Big) + \frac{1}{\sqrt{-g}} \partial_{(\rho}(\sqrt{-g})\partial_{\sigma)} \Big(\frac{\partial f}{\partial \ourB} \Big) = \kappa T_{\rho \sigma} \,.
\end{multline} 

The field equations~(\ref{f(g,b)_final_form}) are symmetric, as they must be, because we varied with respect to the metric. However, one must be very careful now when considering coordinate transformations. The left-hand side clearly contains terms which are not covariant. Moreover, at this point no additional assumptions were made about the matter action and its (non-)invariant properties. The lack off diffeomorphism invariance (up to surface terms) at the level of the total action~(\ref{f(G,B)_action}) plus a matter action implies that we do not automatically automatically have a Bianchi-like identity. As we do not (yet) assume our matter action $S_{\rm matter}[g_{\mu \nu},\Phi]$ to be invariant under diffeomorphisms, we cannot infer the usual covariant conservation $\nabla^{\mu} T_{\mu \nu}=0$. We also note that should $T_{\mu \nu}$ not be a true rank 2 tensor, such an equation would not be well-defined.

Having said this, let us now make the following observation. If we divide~(\ref{f(g,b)_final_form}) by $\partial f/\partial\ourG$ we can isolate the Einstein tensor $G_{\rho\sigma}$ and move all remaining terms to the right-hand side. This would yield an equation of the form
\begin{align} \label{Einstein_tensor_eq}
  G_{\rho\sigma} = \frac{\kappa}{f_{\ourG}} T_{\rho\sigma} + T^{(f)}_{\rho\sigma} \,,
\end{align}
where $f_{\ourG} = \partial f/\partial\ourG$ and $T^{(f)}_{\rho\sigma}$ stands for the collection of all the remaining terms. This formulation has two implications: first, mathematical consistency requires that the right-hand side has to be a rank 2 tensor; second, the twice contracted Bianchi identities imply that the right-hand side must be covariantly conserved
\begin{align}
  \nabla^\rho G_{\rho\sigma} = 0 \qquad \Rightarrow \qquad
  \nabla^\rho \Bigl[\frac{\kappa}{f_{\ourG}} T_{\rho\sigma} +
    T^{(f)}_{\rho\sigma}\Bigr] = 0\,.
\end{align}
Perhaps unexpectedly, we arrived at a conservation equation. Note that we did not assume explicitly the diffeomorphism invariance of the total action. The next section will clarify how this conservation equation emerged.
 
\subsection{Conservation equation and invariance}
\label{subsec:cons}

Let us take a closer look at the diffeomorphisms of action~(\ref{f(G,B)_action}). As previously noted, the Einstein action was in fact zero under diffeomorphisms, owing itself to the fact that the action differed from a coordinate scalar by a boundary term only. Clearly the modified action~(\ref{f(G,B)_action}) does not share this property, so it will not be invariant and it does not differ from a coordinate scalar by a boundary term.

Let us proceed to calculate explicitly how this action transforms under an infinitesimal coordinate transformation. We make use of our previous expressions~(\ref{Lie_G}) and~(\ref{Lie_B}) to find
\begin{align}
  \delta_{\xi} S_{\rm grav}  &= \frac{1}{2\kappa}
  \int \mathcal{L}_{\xi} \big( \sqrt{-g} f(\ourG,\ourB) \big) d^4x
  \nonumber \\ 
  &= \frac{1}{2\kappa}  \int \biggl[\nabla_{\mu} \xi^{\mu} f(\ourG,\ourB) +
  \Bigl(\frac{\partial f(\ourG,\ourB) }{\partial \ourG} \mathcal{L}_{\xi} \ourG +
  \frac{\partial f(\ourG,\ourB) }{\partial \ourB} \mathcal{L}_{\xi} \ourB \Bigr)\biggr]
  \sqrt{-g}\, d^4x \,.
\end{align}
This leads to
\begin{multline}
  \delta_{\xi} S_{\rm grav}=
  \frac{1}{2\kappa} \int
  \biggl[(\partial_{\mu} \xi^{\mu} + \Gamma^{\mu}_{\mu \eta} \xi^{\eta}) f(\ourG,\ourB) +
    \Bigl(\frac{\partial f(\ourG,\ourB)}{\partial \ourG} \partial_{\mu}\ourG \xi^{\mu} +
    \frac{\partial f(\ourG,\ourB) }{\partial \ourB} \partial_{\mu} \ourB \xi^{\mu}  \Bigr) \\
  + M^{\alpha\beta}{}_{\gamma} \partial_\alpha \partial_\beta \xi^\gamma \Big( \frac{\partial f(\ourG,\ourB) }{\partial \ourG} - \frac{\partial f(\ourG,\ourB) }{\partial \ourB} \Big)\biggr] \sqrt{-g}\, d^4x
  \\
   = \textrm{boundary terms} + \frac{1}{2\kappa}  \int M^{\alpha\beta}{}_{\gamma} \partial_\alpha \partial_\beta \xi^\gamma \bigg( \frac{\partial f(\ourG,\ourB) }{\partial \ourG} - \frac{\partial f(\ourG,\ourB) }{\partial \ourB} \bigg) \sqrt{-g}\, d^4x \,.
 \end{multline} 
Discarding the boundary term, and using an abbreviated notation, we are left with 
\begin{equation}
  \delta_{\xi} S_{\rm grav} =   \frac{1}{2\kappa}  \int \sqrt{-g} \big(f_{,\ourG}- f_{,\ourB} \big) M^{\alpha\beta}{}_{\gamma} \partial_\alpha \partial_\beta \xi^{\gamma}\, d^4x \,,
\end{equation}
analogous to Eq.~(\ref{eqn:f0}). As before, we integrate by parts twice and discard the boundary terms, which leads to
\begin{equation}
  \label{f(G,B)_conservation}
  \delta_{\xi} S_{\rm grav} = \frac{1}{2\kappa} \int \partial_{\alpha} \partial_{\beta}
  \Bigl(\sqrt{-g} M^{\alpha\beta}{}_{\gamma} (f_{,\ourG}- f_{,\ourB} ) \Bigr) \xi^{\gamma}\, d^4 x \,.
\end{equation}
If we also include the matter term $\delta_\xi S_{\rm matter}$ as given by~(\ref{eqn:matter}), we have
\begin{align}
  \delta_\xi S_{\rm total} &= \delta_{\xi} S_{\rm grav} + \delta_\xi S_{\rm matter}
  \nonumber \\ &=
  \int \biggl[
    \frac{1}{2\kappa}\partial_{\alpha} \partial_{\beta}
    \Bigl(\sqrt{-g} M^{\alpha\beta}{}_{\gamma} (f_{,\ourG}- f_{,\ourB} ) \Bigr)
    - \nabla_\alpha (\sqrt{-g}T^{\alpha}_{\gamma})\biggr]\xi^{\gamma}\, d^4 x \,.
  \label{eqn:cons0}
\end{align}
If at this point one requires the total action to be invariant under coordinate transformations, one finds a conservation equation, namely the vanishing of the integrand of~(\ref{eqn:cons0}), similar to its GR analogue given by~(\ref{eqn:f2}). One can again regard this equation as a consequence of Noether's theorem. Let us emphasise an important mathematical point here: for the variational formulation to be well-defined, in terms of tensors, we must assume that the integrand of our total action is indeed a well defined scalar, up to possible boundary terms. If the Lagrangian were a pseudo-invariant, as it is for the Einstein action, our action would differ from a true scalar by a boundary term which could be added back into the action to yield a well defined scalar without affecting the resulting field equations.

This conservation equation~(\ref{eqn:cons0}) leads to various interesting points to be addressed. First, let us assume that the matter action is itself based on a true scalar Lagrangian which then implies that the matter energy-momentum tensor is independently conserved $\nabla_\alpha T^{\alpha}_{\gamma} =0$. Requiring $\delta_\xi S_{\rm total} = 0$ then implies the additional equation
\begin{align}
  \partial_{\alpha} \partial_{\beta}
    \Bigl(\sqrt{-g} M^{\alpha\beta}{}_{\gamma} (f_{,\ourG}- f_{,\ourB} )\Bigr) = 0\,,
    \label{eqn:cons1}
\end{align}
which can be satisfied for all possible geometries if $f_{,\ourG}- f_{,\ourB} = 0$. This equation is easily integrated and one finds that all functions of the form $f(\ourG+\ourB)$ are general solutions, assuming a sufficiently regular function $f$. Recalling that this combination is in fact the Ricci scalar, the unique coordinate scalar that can be constructed from $\ourG$ and $\ourB$, we find the expected result that within the family of $f(\ourG,\ourB)$ theories $f(R)$ gravity is the unique diffeomorphism invariant theory. We will show this result explicitly in Section~\ref{sec:f(R)gravity}.

Second, let us also assume that $f(\ourG,\ourB) \neq f(\ourG + \ourB)$. Moreover, we will maintain the assumption that the matter action yields an independent conservation equation $\nabla_\alpha T^{\alpha}_{\gamma} =0$. As before, for the total action to be invariant, Eq.~(\ref{eqn:cons1}) has to be satisfied. In this case it is possible that this term vanishes, however, it will depend both on the metric and the coordinates used. A trivial example would be the situation where $\ourG$ and $\ourB$ are both constants. In this case, and other non-trivial cases, the gravitational action will naively \emph{appear} to be invariant under diffeomorphisms. This situation is similar to that in $f(T)$ gravity when so-called `good tetrads' are considered~\cite{Ferraro:2011us,Tamanini:2012hg}. An explicit example is discussed when considering cosmology in Section~\ref{sec:fgcosmology}.

Lastly, if we drop the assumption of the matter conservation equation, only the integrand of~(\ref{f(G,B)_conservation}) has to vanish to achieve diffeomorphism invariance of the total action. This would allow us to study theories in which for instance the dark matter energy density decreases over time while the dark energy density, here modelled through the contributions of $f(\ourG,\ourB)$, can increase over time.

\subsection{Retrieving \texorpdfstring{$f(R)$}{f(R)} gravity}
\label{sec:f(R)gravity}

If we take $f(\ourG,\ourB)$ to be a function of the Ricci scalar only, that is $f(\ourG,\ourB) = f(\ourG + \ourB) = f(R)$, we readily recover the $f(R)$ field equations. First we use that
\begin{equation}
    \frac{\partial f(\ourG+\ourB)}{\partial \ourG} = \frac{\partial f(\ourG+\ourB)}{\partial \ourB} \ = \frac{\partial f(R)}{\partial R} \,.
\end{equation}
With this simplification we can rewrite the left-hand side of equation~(\ref{f(g,b)_final_form}) as follows
\begin{multline}
  \label{f(GB)variation_expansion1}
  \frac{\partial f}{\partial R} \Big[ G_{\rho \sigma} + \frac{1}{2} g_{\rho \sigma} \ourG +
    \frac{1}{2} g_{\rho \sigma} \ourB \Big] - \frac{1}{2} g_{\rho \sigma} f(R) +
  g_{\rho \sigma} \partial^{\mu} \partial_{\mu} \Big(\frac{\partial f}{\partial R} \Big) -
  \partial_{\rho} \partial_{\sigma} \Big( \frac{\partial f}{\partial R} \Big) 
  \\
  + \partial_{\gamma} \Big( \frac{\partial f}{\partial R} \Big)
  \Big[
    \frac{1}{2} E_{\rho \sigma}{}^{\gamma} + \frac{1}{2} g_{\rho \sigma} \partial_{\mu}g^{\mu \gamma} +
    \frac{1}{\sqrt{-g}} \partial_{(\rho}(\sqrt{-g}) \delta^{\gamma}_{\sigma)}  
    \Big] \,.
\end{multline}
Expanding the definition of $E_{\rho \sigma}{}^{\gamma}$~(\ref{E}), writing the partial derivatives of the metric in terms of Christoffel symbols and using $\ourG + \ourB = R$ gives
\begin{multline}
  \label{f(GB)variation_expansion2}
    \frac{\partial f}{\partial R} \Big[ G_{\rho \sigma} + \frac{1}{2} g_{\rho \sigma}R \Big]
    - \frac{1}{2} g_{\rho \sigma} f(R) + g_{\rho \sigma} \partial^{\mu} \partial_{\mu} \Big(\frac{\partial f}{\partial R} \Big)  - \partial_{\rho} \partial_{\sigma} \Big( \frac{\partial f}{\partial R} \Big) 
    \\
    + \partial_{\gamma} \Big( \frac{\partial f}{\partial R} \Big)
    \Big[ \Gamma^{\gamma}_{\rho \sigma} - g_{\rho \sigma} g^{\mu \nu} \Gamma^{\gamma}_{\mu \nu}
    \Big]  \,.
\end{multline}
The term in the first square bracket is the Ricci tensor $R_{\rho \sigma}$. The other connection and partial derivative terms are simply the covariant derivatives acting on $\partial f/\partial R$, which, unlike $\ourG$ and $\ourB$, is a true coordinate scalar. To see this more clearly, begin with the $f(R)$ field equations \cite{Sotiriou:2008rp,DeFelice:2010aj} given by
\begin{equation}
  \label{f(R)field_equation}
    \frac{\partial f}{\partial R} R_{\rho \sigma} + \big[g_{\rho \sigma} \Box - \nabla_{\rho} \nabla_{\sigma}\big]\frac{\partial f}{\partial R} - \frac{1}{2} g_{\rho \sigma} f(R) \,,
\end{equation}
and expand the derivative terms
\begin{multline}
    \big[g_{\rho \sigma} \Box - \nabla_{\rho} \nabla_{\sigma}\big] f_{,R} =
    g_{\rho \sigma} g^{\mu \nu} \big(\partial_{\mu} \nabla_{\nu} - \Gamma^{\gamma}_{\mu \nu} \nabla_{\gamma} \big) f_{,R} - 
     \nabla_{\rho} \partial_{\sigma} f_{,R} \\
    = \Big( g_{\rho \sigma} g^{\mu \nu} \partial_{\mu} \partial_{\nu} 
    - g_{\rho \sigma} g^{\mu \nu} \Gamma^{\gamma}_{\mu \nu} \partial_{\gamma} - \partial_{\rho} \partial_{\sigma} + \Gamma^{\gamma}_{\rho \sigma} \partial_{\gamma} \Big) f_{,R} \,.
\end{multline}
Comparison of (\ref{f(GB)variation_expansion2}) with the expressions above shows that these are indeed the same, such that $f(\ourG+\ourB)$ yields the standard $f(R)$ field equations. These are generally fourth order, due to the second order terms in the Ricci scalar (which are confined to the boundary in GR) contributing to the equations of motion when $f(R)$ is a non-linear function.

\subsection{\texorpdfstring{$f(\ourG)$}{f(G)} gravity}

Theories of the form $f(\ourG)$, i.e.~modifications of the Einstein action~(\ref{Einstein_Action}),  will produce second order equations which are (in general) non-covariant when $f(\ourG) \neq c_1 \ourG$. Again, the $f(\ourG)$ field equations will depend on the coordinate system used. This means that solutions cannot generally be transformed into solutions in other coordinates, but we can find `preferred' sets of coordinates which lead to interesting results: i.e.~dynamics that differ from GR.

This approach is very similar to $f(Q)$ gravity~\cite{BeltranJimenez:2017tkd} where close links between the non-metricity scalar $Q$ and $\ourG$ were already explored. The non-metricity scalar depends on the first partial derivatives of the metric and hence also yields second order field equations. The difference between both approaches will be discussed in greater detail when boundary terms will be taken into account.

Interestingly, this approach was recently considered in~\cite{Milgrom:2019rtd} where it was motivated by modified Newtonian dynamics (MOND) and showed interesting links with $f(Q)$ gravity. It was argued that a gravitational theory might not be diffeomorphism invariant in the limit of small accelerations where a modified theory can be employed to explain flattened galactic rotation curves. 

The $f(\ourG)$ field equations follow directly from~(\ref{f(g,b)_final_form}) and are given by
\begin{align}
  \label{f(G)field_equation}
  f'(\ourG) \Big[G_{\rho \sigma} + \frac{1}{2} g_{\rho \sigma} \ourG \Big] + \frac{1}{2} f''(\ourG) E_{\rho \sigma}{}^{\gamma} \partial_{\gamma} \ourG - \frac{1}{2} g_{\rho \sigma} f(\ourG) =
  \kappa T_{\rho \sigma} \,.
\end{align}
One can immediately see that in this compact form, where the Christoffel symbols are contained within the object $E_{\rho \sigma}{}^{\gamma}$, the field equations resemble those found in $f(T)$ gravity (in its covariant form~\cite{Li:2011wu}). Two particular examples where the $f(T)$ and $f(\ourG)$ field equations are identical are the Friedmann–Lema{\^{\i}}tre–Robertson–Walker (FLRW) metric in Cartesian coordinates and the Schwarzschild metric in Cartesian isotropic coordinates, we will show these equations explicitly below.

The integrand of the conservation equation~(\ref{eqn:cons0}) becomes
\begin{align}
  \label{f(G)_conservation}
  \frac{1}{2\kappa}\partial_{\alpha} \partial_{\beta}
  \bigl(\sqrt{-g} M^{\alpha\beta}{}_{\gamma} f'(\ourG)\bigr) - \nabla_\alpha T^\alpha_\gamma \sqrt{-g} = 0 \,.
\end{align}
Let us rewrite the field equations~(\ref{f(G)field_equation}) into an Einstein-like form. We divide by $f'$ and isolate the Einstein tensor, which yields
\begin{align} \label{f(G)_2}
  G_{\rho\sigma} = \frac{\kappa}{f'} T_{\rho\sigma} + \frac{1}{2}\frac{1}{f'} \Bigl(
  (f - f' \ourG) g_{\rho\sigma} -
  f'' E_{\rho \sigma}{}^{\gamma} \partial_{\gamma} \ourG \Bigr) \,.
\end{align}
The second term of the right-hand side can be interpreted as the effective energy-momentum tensor of the modified theory of gravity, similar to the setup in $f(R)$ gravity. Since the covariant trace of the Einstein tensor vanishes, one can again derive a conservation equation which will be equivalent to~(\ref{f(G)_conservation}). It is interesting to note that it is not clear from first principles how the covariant derivative should act on the quantities $E$ or $\ourG$ as these are a non-tensorial rank 3 object and pseudo scalar, respectively. The advantage of~(\ref{f(G)_conservation}) is that its formulation only uses partial derivatives which are well-defined on all objects, tensorial and non-tensorial. In this sense one could combine these two equations to read off the `meaning' of these covariant derivatives of $E$ and $\ourG$. However, we will not require this for what follows.

We can think of $f(R)$ gravity as the unique theory based on a true scalar while $f(\ourG)$ gravity can be seen as the unique theory with second order equations within the $f(\ourG,\ourB)$ family. This is visualised in Fig.~\ref{fig1}. This can be seen as the `coordinate' analogue of the discussion in~\cite{Bahamonde:2015zma} in the context of modified teleparallel theories of gravity. Within this family of theories, General Relativity emerges as the unique diffeomorphism invariant theory that has second order field equations.

\begin{figure}[!hbt]
\centering
\begin{tikzpicture}
  \matrix (m) [matrix of math nodes,row sep=6em,column sep=10em,minimum width=2em]
          {f(\ourG,\ourB) & f(\ourG) \\
            f(\ourG+\ourB) & \text{GR} \\};
  \path[-stealth]
  (m-1-1) edge node [above] {second order} (m-1-2)
  (m-1-1) edge node [left] {diff.~invariant} (m-2-1)
  (m-2-1) edge node {} (m-2-2)
  (m-1-2) edge node {} (m-2-2);
\end{tikzpicture}
\caption{Relationship between $f(\ourG,\ourB)$ gravity and General Relativity.}
\label{fig1}
\end{figure}

The discussion of~\cite{Milgrom:2019rtd} also restricted the form of the function $f(\ourG)$ in order to yield GR in a suitable limit, $f(\ourG) \rightarrow \ourG + c_1$ for large $\ourG$ and $c_1$ being a constant. The so-called deep MOND limit is attained when $f(\ourG) \rightarrow \alpha \ourG^{3/2} + c_2$ for small $\ourG$. 

\subsection{\texorpdfstring{$f(\ourG)$}{f(G)} cosmology}
\label{sec:fgcosmology}

Comparison of the cosmological equation with those in $f(T)$ and $f(Q)$ gravity shows that theses are in fact identical. At first sight this result is surprising and one is tempted to attribute it to the symmetry of the FLRW spacetime. In the following we will investigate the relations between these theories is greater detail. 

For a homogeneous and isotropic cosmological spacetime given by the spatially flat FLRW metric in Cartesian coordinates the line element is given by
\begin{align}
ds^2 = - N(t)^2 dt^2 + a(t)^2 \bigl(dx^2 + dy^2 + dz^2 \bigr) \,.
\end{align}
We assume the energy-momentum tensor to be modelled by a single perfect fluid with pressure $p$ and energy density $\rho$. We will require an equation of state to close the system of equations if closed form solutions are sought. The $f(\ourG)$ field equations are
\begin{align}
  \label{f(G)FLRW_Hub1}
  f(\ourG) N^2 + 12 H^2 f'(\ourG)
  &= 2  \kappa \rho \,, \\
  \label{f(G)FLRW_Hub2}
  -f(\ourG) - f'(\ourG) \frac{4}{N^2} \Bigl(\dot{H} + 3H^2 - H\frac{\dot{N}}{N} \Bigr) 
  + f''(\ourG) \frac{48H^2}{N^4} \Bigl(\dot{H}-H\frac{\dot{N}}{N}\Bigr)
  &= 2  \kappa p \,,
\end{align}
where $H = \dot{a}/a$ is the Hubble function. In these coordinates $\ourG$ is related to the Hubble parameter by
\begin{align}
  \label{FLRW_G}
  \ourG = -6\frac{\dot{a}^2}{a^2 N^2} = -6\frac{H^2}{N^2} \,.
\end{align}
Note that these equations satisfy the continuity equation. This can also be seen by noting that the conservation equation~(\ref{f(G)_conservation}) vanishes for this choice of metric and coordinates. If one works in the full $f(\ourG,\ourB)$ theory the conservation equation~(\ref{f(G,B)_conservation}) is also zero. This corresponds to the second possibility discussed in Section~\ref{subsec:cons} where the equations appear to be invariant due to the specific choice of coordinates. 

The above cosmological equations~(\ref{f(G)FLRW_Hub1}) and~(\ref{f(G)FLRW_Hub2}) are identical in form to the $f(T)$ cosmological field equations, assuming $N(t)=1$, with the torsion scalar $T$ replaced with our bulk term $\ourG$ (e.g.~see equations (9) and (10) in \cite{Bengochea:2008gz}). Similarly, the torsion scalar $T$ is equal to minus the bulk term $\ourG$. Hence the dynamics for both theories, given that coordinates and tetrads are chosen appropriately, are equivalent, see also~\cite{Coley:2019zld} for recent discussion in the context of teleparallel theories. In $f(Q)$ gravity it is known that the cosmological field equations are also of the same form as those found in $f(T)$ gravity \cite{Jimenez:2019ovq,Frusciante:2021sio}, and therefore the background dynamics are again equivalent. In this case $Q = 6H^2 = -\ourG$ with $N(t)=1$ and a slight redefinition of the function $f$ yields equivalence. After briefly stating the spherically symmetric field equations we will proceed to discuss this equivalence thoroughly in Section~\ref{sec:equiv}.

\subsection{\texorpdfstring{$f(\ourG)$}{f(G)} spherical symmetry}
\label{subsec:sphere}

Next we consider a spherically symmetric vacuum spacetime with isotropic coordinates
\begin{align} 
  ds^2 = -{A(R)}^2 dt^2 + {B(R)}^2 \bigl(dx^2 + dy^2 + dz^2 \bigr)\,, \qquad 
  R = \sqrt{x^2 + y^2 + z^2} \,.
\end{align}
For the given metric the $f(\ourG)$ field equations~(\ref{f(G)field_equation}) yield two independent equations, which can be written as
\begin{align}
  \frac{f(\ourG)}{4} - \Bigl(\frac{A'}{A}\frac{B'}{B^3}+\frac{2}{R}\frac{B'}{B^3}+\frac{B''}{B^3}\Bigr)f'(\ourG)-
  \frac{B'}{B^3}\ourG' f''(\ourG) &= 0 \,, \\[1ex]
  \frac{B \left(2 R A' B'+A \left(B'-R B''\right)\right)+B^2
    \left(A'-R A''\right)+2 R A B'^2}{R B \left(B A'+A B'\right)} -
  \ourG'\frac{f''(\ourG)}{f'(\ourG)} &= 0 \,,
\end{align}
where we've used the shorthand $A' = dA/dR$. In these coordinates $\ourG$ is
\begin{align}
  \label{G_Isotropic}
  \ourG=2\frac{B'}{B^3}\Bigl(2\frac{A'}{A} + \frac{B'}{B}\Bigr) \,.
\end{align}
Just as in the previous example, the $f(T)$ field equations are identical to those presented here and the torsion scalar $T$ is again equal to minus~(\ref{G_Isotropic}). It is important to emphasise that these results only hold for the chosen coordinates and tetrads. These two nontrivial examples motivate us to look closer at the relation between the seemingly different formulations of modified gravity. Also note that this choice of metric and coordinates again satisfies our conservation equation in both the $f(\ourG)$ case~(\ref{f(G)_conservation}) and the $f(\ourG,\ourB)$ case~(\ref{f(G,B)_conservation}). When using the standard Schwarzschild coordinates, one finds an undesirable off-diagonal field equation proportional to $f''(\ourG) = 0$, similar to the situation in $f(T)$ gravity where it is rather challenging to identity a suitable static and spherically symmetric tetrad field~\cite{Ferraro:2011ks}.

\subsection{A first glimpse at equivalence}

The field equations for the three modified theories, $f(\ourG)$, $f(T)$ and $f(Q)$, can be stated in a similar form as follows
\begin{align} 
    f'(\ourG) \Big[G_{\rho \sigma} + \frac{1}{2} g_{\rho \sigma} \ourG \Big] + \frac{1}{2} f''(\ourG) E_{\rho \sigma}{}^{\gamma} \partial_{\gamma} \ourG - \frac{1}{2} g_{\rho \sigma} f(\ourG) &= \kappa T_{\rho \sigma} \,, \\
    f'(T) \Big[G_{\rho \sigma} - \frac{1}{2} g_{\rho \sigma} T \Big] + f''(T) \mathscr{S}_{\rho \sigma}{}^{\gamma} \partial_{\gamma} T + \frac{1}{2} g_{\rho \sigma} f(T) &= \kappa \Theta_{\rho \sigma} \,, \\
    f'(Q) \Big[G_{\rho \sigma} - \frac{1}{2} g_{\rho \sigma} Q \Big] + 2 f''(Q) P^{\lambda}{}_{\rho \sigma} \partial_{\lambda} Q + \frac{1}{2} g_{\rho \sigma} f(Q) &= \kappa T_{\rho \sigma} \,,
\end{align}
see \cite{Li:2011wu} and \cite{Jimenez:2019ovq,Harko:2018gxr} for $f(T)$ and $f(Q)$, respectively. The objects $E^{\mu \nu \rho}$, $\mathscr{S}^{\mu \nu \rho}$ and $P^{\mu \nu \rho}$ act as the superpotential of each theory. It has already been noted that in the context of $f(Q)$ gravity, whilst working in the so-called coincidence gauge, the non-metricity scalar $Q$ reduces to $-\ourG$~\cite{BeltranJimenez:2017tkd}. In the context of the symmetric teleparallel equivalent of general relativity (STEGR) the action is then just the Einstein action and leads to the Einstein field equations, see for instance~\cite{MOLLER1958347,Nester:1998mp,TP:2010}. One also finds that in the coincidence gauge $P^{\lambda}{}_{\mu \nu} = \frac{1}{4} E_{\mu \nu}{}^{\lambda}$, in which case the $f(Q)$ field equations are identical to the $f(\ourG)$ equations as one would expect at this point. However, the geometric setting of all three theories differs, which we will explore more in the following. 

Clearly, the above equivalence only holds provided one works with appropriate coordinates or with an appropriate tetrad. It is well known in $f(T)$ gravity that some tetrads yield undesirable field equations which generally imply $f''(T)=0$, thereby reducing the model back to GR\@. It is also worth emphasising at this point that in TEGR and its generalisations, the tetrad field $e^a_\mu$ becomes the dynamical variable while the metric takes a secondary role. This also means one deals with 16 field equations instead of 10.

\section{Towards equivalence}
\label{sec:equiv}

\subsection{Affine connection}

The starting point of teleparallel gravity is the well-known relation $R=-T+B$, while the starting point for the symmetric teleparallel formulation of General Relativity is of similar structure. Let us now consider spaces with curvature, torsion and non-metricity (a general affine space) so that the affine connection $\bar{\Gamma}$ can be written symbolically as $\bar{\Gamma} = \Gamma + K$ where $K$ is the contortion tensor that represents the difference between the Levi-Civita  (or Christoffel) connection $\Gamma$ and the full connection $\bar{\Gamma}$ \cite{JS1954}. This term consists of permutations of the torsion and non-metricity tensors
\begin{align}
  K_{\mu\lambda}{}^{\kappa} &= g^{\kappa\rho}
  \bigl(-T_{\{\mu\rho\lambda\}} + \frac{1}{2}Q_{\{\mu\rho\lambda\}}\bigr) \,,\\
  T_{\mu\nu}{}^\kappa &= \frac{1}{2}\bigl(\bar{\Gamma}^{\kappa}_{\mu\nu} -
  \bar{\Gamma}^{\kappa}_{\nu\mu}\bigr)\,, \qquad
  Q_{\mu}{}^{\kappa\lambda} = \bar{\nabla}_\mu g^{\kappa\lambda} \,.
\end{align}
Here $\bar{\nabla}$ stands for the covariant derivative using the affine connection. The Riemann tensor for the affine connection is given by
\begin{align}
\bar{R}_{\mu \nu \lambda}{}^{\rho} = 2 \bar{\Gamma}_{\kappa [\mu}^{\rho} \bar{\Gamma}^{\kappa}_{\nu] \lambda}+ 2 \partial_{[\mu} \bar{\Gamma}_{\nu]\lambda}^{\rho} \,,
\end{align}
and the complete Ricci scalar is given by 
\begin{align}
  \overline{R} = R + \nabla_\kappa K_{\mu}{}^{\mu\kappa} - \nabla_\kappa K_{\mu}{}^{\kappa \mu} +
  K_{\kappa\rho}{}^\kappa K^\lambda{}_\lambda{}^\rho - K^\lambda{}_\rho{}^\kappa K_{\kappa\lambda}{}^\rho \,,
\end{align}
where $\nabla$ stands for the covariant derivative using the Levi-Civita connection. Using that the Ricci scalar $R$ can be written as in~(\ref{EH_G_B_Action}), we write the previous relation as
\begin{align}
  \bar{R} = \ourG + \ourB + \nabla_\kappa K_{\mu}{}^{\mu\kappa} - \nabla_\kappa K_{\mu}{}^{\kappa\mu} +
  K_{\kappa\rho}{}^\kappa K^\lambda{}_\lambda{}^\rho - K^\lambda{}_\rho{}^\kappa K_{\kappa\lambda}{}^\rho \,.
  \label{equiv1}
\end{align}
We note that the third and fourth terms on the right-hand side are also boundary terms as the covariant derivative acts on rank-one tensors due to the summations.

In TEGR one works with a flat manifold where the Riemann curvature tensor of the complete connection vanishes, and it is assumed that non-metricity vanishes identically. This means Eq.~(\ref{equiv1}) becomes
\begin{align}
  0 = \ourG + \ourB - 2 \nabla_\kappa K_{\mu}{}^{\kappa\mu} +
  K_{\kappa\rho}{}^\kappa K^\lambda{}_\lambda{}^\rho - K^\lambda{}_\rho{}^\kappa K_{\kappa\lambda}{}^\rho \,,
  \label{equiv2}
\end{align}
where now the contortion terms only depend on the torsion tensor. This is often written in the form
\begin{align}
  0 = \ourG + \ourB + T - B_{T} \,.
  \label{equiv3}
\end{align}
Here $T$ is the torsion scalar
\begin{align}
  T = T^{\rho\lambda\kappa} T_{\rho\lambda\kappa} + 2 T^{\kappa\lambda\rho}T_{\kappa\rho\lambda} - 4
  T_{\rho}{}^{\kappa}{}_{\kappa} T^{\rho\lambda}{}_{\lambda} \,,
\end{align}  
and $B_T$ is its boundary term, such that $-T + B_T = R$. As stated before, we follow the conventions of~\cite{JS1954}. The superpotential $\mathscr{S}^{\kappa\rho\lambda}$ is defined through the relation $T = \mathscr{S}^{\kappa\rho\lambda} T_{\rho\lambda\kappa}$ and is given by
\begin{align}
  \mathscr{S}^{\kappa}{}_{\rho\lambda} = T_{\rho\lambda}{}^{\kappa} - T^{\kappa}{}_{\rho\lambda} +
  T^{\kappa}{}_{\lambda\rho} + 2 (\delta^\kappa_\rho T_{\lambda}{}^{\sigma}{}_{\sigma} -
  \delta^\kappa_\lambda T_{\rho}{}^{\sigma}{}_{\sigma}) \,.
\end{align}
It should be noted that different conventions used in the literature can result in slightly different pre-factors, these mainly go back to the definition of torsion.

In the so called symmetric teleparallel gravity models one again works on a manifold with vanishing curvature, however, with non-metricity instead of torsion. In this case, Eq.~(\ref{equiv1}) becomes 
\begin{align}
  \label{equiv8}
  0 = \ourG + \ourB + Q + B_{Q} \,,
\end{align}
where $Q$ is the non-metricity scalar and $B_Q$ its boundary term. Again, we can write the Ricci scalar as $Q + B_Q = -R$. Combining these three settings into one yields the fundamental equation
\begin{align}
  \label{fund1}
  \overline{R} = \ourG + \ourB + T - B_{T} + Q + B_Q + \ourC\,,
\end{align}
where $\ourC$ contains the torsion--non-metricity cross terms which appear in~(\ref{equiv1}). These cross terms can be conveniently written using the superpotential
\begin{align}
  \label{cross1}
  \ourC = \mathscr{S}^{\kappa\rho\lambda} Q_{\lambda\kappa\rho} =
  2 \bigl(
  Q_{\kappa\rho\lambda} T^{\lambda\kappa\rho} + Q_{\rho}{}^{\sigma}{}_{\sigma} T^{\rho\kappa}{}_{\kappa} -
  Q^{\sigma}{}_{\sigma\rho} T^{\rho\kappa}{}_{\kappa}
  \bigr) \,,
\end{align}
see for instance~\cite{Jimenez:2019ghw}. The identity~(\ref{fund1}) is the starting point for our subsequent discussion.

\subsection{Boundary terms and TEGR}

If equivalence between different theories can be established, it will be important to understand how the different boundary terms are related to each other. Clearly, the geometrical setting that underlies $f(T)$ gravity, with its use of tetrads as the basic dynamical variable, appears difficult to be reconciled with the metric formulation or the non-metricity formulation. First we will show that the boundary term appearing in $f(T)$ gravity $B_T$ differs from our boundary term $\ourB$ by yet another boundary term which we will call $b_T$.

Starting from the previous equation~(\ref{equiv3}) $0 = \ourG + \ourB + T - B_{T}$ the boundary term $B_{T}$ is explicitly given by
\begin{align}
  B_T = \frac{2}{e}\partial_\mu(e T^\mu) = 2 \nabla_\mu T^\mu \,, \qquad
  T^\mu = T^{\mu}{}_\lambda{}^\lambda \,.
  \label{equiv4}
\end{align}
We follow the standard formulation of TEGR where the spin-connection is set to zero $\omega_\mu{}^a{}_b = 0$, the torsion tensor is explicitly given in terms of the tetrad field only and one finds the following expression for the boundary term
\begin{align}
  B_T = \frac{2}{e}\partial_\mu\Bigl(e g^{\mu\nu} E^\lambda_a (\partial_\lambda e_\nu^a -
  \partial_\nu e_\lambda^a)\Bigr) \,.
  \label{equiv4a}
\end{align}
We will now look at our original boundary term $\ourB$ re-expressed using tetrads. Recall definition~(\ref{B_boundary}) which reads
\begin{align}
  \ourB = \frac{1}{\sqrt{-g}}\partial_\nu\Bigl(\frac{1}{\sqrt{-g}}\partial_\mu(g g^{\mu\nu})\Bigr) =
  -\frac{1}{\sqrt{-g}} \partial_\nu \Bigl(\sqrt{-g}(g^{\mu\nu} g^{\beta\alpha} -
  g^{\mu\alpha} g^{\nu\beta})\partial_\mu g_{\alpha\beta}\Bigr) \,,
  \label{equiv5}
\end{align}
where we have used the standard relations
\begin{align}
  \partial_{\mu} g^{\sigma \rho} = - g^{\sigma \alpha} g^{\rho\beta} \partial_{\mu} g_{\alpha\beta}\,,
  \qquad
  \partial_\nu \sqrt{-g} = -\frac{1}{2}\sqrt{-g} g_{\alpha\beta} \partial_\nu g^{\alpha\beta}\,,
\end{align}
etc. Rewriting the metric tensor in terms of tetrads using $g^{\alpha\beta} = E^\alpha_m E^\beta_n \eta^{mn}$ and $e=\sqrt{-g}$ we find
\begin{align}
  \ourB =
  -\frac{1}{e}\partial_\nu\Bigl(e
  (g^{\mu\nu} g^{\beta\alpha} - g^{\mu\alpha} g^{\nu\beta})e_\alpha^m \partial_\mu e_\beta^n +
  (g^{\mu\nu} g^{\beta\alpha} - g^{\mu\alpha} g^{\nu\beta})e_\beta^n \partial_\mu e_\alpha^m)
  \eta_{mn}\Bigr) \,.
  \label{equiv6_bb}
\end{align}
After expanding this out and simplifying terms we arrive at
\begin{align}
  \ourB =
  \frac{1}{e}\partial_\mu\Bigl(e
  g^{\mu\nu} E^\lambda_a(\partial_\lambda e_\nu^a - \partial_\nu e_\lambda^a) +
  e g^{\beta\nu} E^\lambda_a(
  \delta_\lambda^\mu \partial_\beta e_\nu^a -
  \delta^\mu_\beta \partial_\nu e_\lambda^a)\Bigr) \,,
  \label{equiv6_bb2}
\end{align}
which is a lengthy calculation and constitutes the first key result of this part. Comparison of the first two terms in~(\ref{equiv6_bb2}) with~(\ref{equiv4a}) shows that the boundary term $\ourB$ contains the teleparallel boundary term $B_T$
\begin{align}
  \ourB = \frac{1}{2} B_T + \frac{1}{e}\partial_\mu
  \Bigl(e g^{\beta\nu} E^\lambda_a(\delta_\lambda^\mu \partial_\beta e_\nu^a -
  \delta^\mu_\beta \partial_\nu e_\lambda^a)\Bigr) =:
  \frac{1}{2} B_T + \frac{1}{2} B_T^{(2)} \,.
\end{align}
In other words, this result states that both boundary terms differ by yet another boundary term, which we will denote by $B_T^{(2)}$. Due to the many different boundary terms notation can become slightly difficult but we will try to keep the notation as clear as possible.

\subsection{A new boundary and TEGR equivalence}
\label{subsec:tegr}

It is of interest to consider the following linear combination of the previous boundary terms. Let us define the term
\begin{align}
  b_T := \ourB - B_T = \frac{1}{2} B_T^{(2)} - \frac{1}{2} B_T =
  \frac{1}{e} \partial_{\nu} \Big[e g^{\mu \alpha} E^{\nu}_{a} \partial_{\mu} e^{a}_{\alpha} -e g^{\nu \alpha} E^{\mu}_{a} \partial_{\mu} e^{a}_{\alpha} \Big] \,.
  \label{def:bt}
\end{align}
Using the fact that the square bracket is skew-symmetric in $\mu \nu$, we can show that the term, in fact, does not depend on any second order derivatives
\begin{align} \label{Bnew}
  b_T &= \frac{1}{e} \partial_{\nu} \Big[e g^{\mu \alpha} E^{\nu}_{a} \partial_{\mu} e^{a}_{\alpha} - e g^{\nu \alpha} E^{\mu}_{a} \partial_{\mu} e^{a}_{\alpha} \Big] =
  \frac{2}{e} \partial_{\nu} \Big[e \partial_{\mu} e^{a}_{\alpha}
    \big(g^{\alpha [ \mu} E^{\nu ]}_{a} \big) \Big]
  \nonumber \\ &=
  \frac{2}{e} \partial_{\mu}\partial_{\nu} e^{a}_{\alpha}
  \big(e g^{\alpha [ \mu} E^{\nu ]}_{a} \big) + \frac{2}{e} (\partial_{\mu}e^{a}_{\alpha}) \partial_{\nu}
  \big(e g^{\alpha [ \mu} E^{\nu ]}_{a} \big)
  \nonumber \\
&= \frac{2}{e} (\partial_{\mu}e^{a}_{\alpha}) \partial_{\nu} \big(e g^{\alpha [ \mu} E^{\nu ]}_{a} \big)  \,,
\end{align}
where in the second last line we have expanded using the product rule and then noted that the first term vanishes. The partial derivatives are symmetric whilst the bracket is skew-symmetric. Going back once more to our fundamental equation $0 = \ourG + \ourB + T - B_{T}$ we can now write
\begin{align}
  \ourG + \ourB + T - B_{T} &= \ourG + \frac{1}{2} B_T + \frac{1}{2} B_T^{(2)} + T - B_{T} =
  \ourG + T + \frac{1}{2} B_T^{(2)} - \frac{1}{2} B_T \,.
\end{align}
This identity establishes the first result of this section
\begin{align}
  \ourG + T + b_T = 0 \qquad \Rightarrow \qquad \ourG &= -T - b_T \,.
  \label{G_equiv}
\end{align}

From this final relation~(\ref{G_equiv}) it is clear that $b_T$ must be first order and quadratic in derivatives of the tetrad, because both $\ourG$ and $T$ are. This is verified in the final form of equation~(\ref{Bnew}). Another property of this new boundary term is that it vanishes for any diagonal tetrad. In this case equation~(\ref{G_equiv}) simply reduces to $\ourG = - T$.

Next, we note that $b_T$ is neither a Lorentz scalar nor a coordinate scalar! This boundary term encodes information describing how the torsion scalar $T$ fails to be Lorentz invariant. Equivalently, it measures how the bulk term $\ourG$ fails to be invariant under coordinate transformations. It therefore contains important information about the structure of the spacetime and the tangent space for a given tetrad field. In order for the field equations of $f(\ourG)$ and $f(T)$ gravity to match, it is necessary but not sufficient for $b_T$ to vanish. The coordinates and tetrads chosen in Sections~\ref{sec:fgcosmology} \&~\ref{subsec:sphere} satisfy this very condition.

If one wishes to work in the modified teleparallel setting where the spin-connection is set to zero, as in the above, then it is well known that the theory is not invariant under local Lorentz transformations~\cite{Li:2010cg,Capozziello:2011et,Cai:2015emx}. Any attempt at partially restoring the Lorentz symmetry, without altering the spin connection, which means by considering only a restricted class of tetrads must enforce the condition $b_T=0$. This term hence provides us with a new approach to understand the so-called remnant symmetry of $f(T)$ gravity~\cite{Chen:2014qtl,Ferraro:2014owa}. It should be noted, however, that there exists an alternative approach to modified teleparallel theories of gravity, which abandons the requirement of having a vanishing spin connection, thereby allowing for a covariant formulation of the theory~\cite{Krssak:2018ywd}. For completeness we should state that this approach is not free of criticism~\cite{Bejarano:2019fii,Maluf:2018coz} though.

\subsection{Boundary term of symmetric teleparallel gravity}

The ideas and methods of the previous sections can also be applied to symmetric teleparallel gravity as we will now show. Again one works with a flat manifold where the Riemann curvature tensor of the complete connection vanishes, however, in this case non-metricity is present but torsion is assumed to vanish identically. As before, we have~(\ref{equiv2}) which now becomes
\begin{align}
  0 = \ourG + \ourB + Q + B_Q \,.
  \label{equiv8a}
\end{align}
Here the non-metricity scalar $Q$ is a quadratic combination of the non-metricity tensor, analogous to the torsion scalar, while the boundary term $B_Q$ is
\begin{align}
  B_Q = \nabla_\kappa K_{\mu}{}^{\mu\kappa} - \nabla_\kappa K_{\mu}{}^{\kappa\mu} \,.
  \label{equiv9}
\end{align}
Writing out the contortion explicitly in terms of non-metricity, see~\cite{JS1954}, gives
\begin{align}
  K_\mu{}^{\lambda\kappa} = \frac{1}{2}\bigl(
  Q_\mu{}^{\kappa\lambda}-Q^{\kappa\lambda}{}_\mu + Q^\lambda{}_\mu{}^\kappa\bigr) \,,
  \label{equiv10}
\end{align}
from which we find the two contractions
\begin{align}
  K_\mu{}^{\mu\kappa} = \frac{1}{2}\bigl(2 Q_\mu{}^{\mu\kappa}-Q^{\kappa\mu}{}_\mu\bigr) \,, \qquad
  K_\mu{}^{\kappa\mu} = \frac{1}{2} Q^{\kappa}{}_{\mu}{}^\mu \,.
  \label{equiv11}
\end{align}
Therefore, the boundary term can then be written as
\begin{align}
  B_Q = \nabla_\kappa K_\mu{}^{\mu\kappa} - \nabla_\kappa  K_\mu{}^{\kappa\mu} =
  \nabla_\kappa \bigl(Q_\mu{}^{\mu\kappa} - Q^{\kappa\mu}{}_\mu\bigr) =
  \frac{1}{\sqrt{-g}}\partial_\kappa \bigl(\sqrt{-g}(Q_\mu{}^{\mu\kappa} - Q^{\kappa\mu}{}_\mu)\bigr) \,.
  \label{equiv12}
\end{align}

Making the covariant derivative explicit on the metric tensor, the non-metricity tensor can be written as
\begin{align}
  \label{non-metricity}
  Q_{\lambda \mu \nu} = -\bar{\nabla}_{\lambda} g_{\mu \nu} =
  -\partial_{\lambda} g_{\mu \nu} + 2\bar{\Gamma}^{\kappa}_{\lambda (\mu} g_{\nu) \kappa} \,,
\end{align}
where $\bar{\Gamma}$, as before, is the affine connection. Starting with an affine connection which vanishes and then making a coordinate transformation gives
\begin{align}
  \label{nonmetricity_connection}
  \bar{\Gamma}^{\lambda}_{\mu \nu} =
  \frac{\partial x^{\lambda}}{\partial \xi^{\sigma}} \partial_{\mu}\partial_{\nu} \xi^{\sigma} \,,
\end{align}
where $\xi^{\sigma} = \xi^{\sigma} (x^{\mu})$ are functions of the coordinates $x^{\mu}$. It is common in $f(Q)$ gravity to work in the so-called \emph{coincident gauge}, see~\cite{BeltranJimenez:2017tkd}, where the coordinates are chosen such that the connection~(\ref{nonmetricity_connection}) vanishes. This is often denoted as $\accentset{\circ}{\Gamma} = 0$. Then the non-metricity tensor~(\ref{non-metricity}) is determined by the partial derivatives of the metric
\begin{align}
  \accentset{\circ}{Q}_{\lambda \mu \nu} =  -\partial_{\lambda} g_{\mu \nu} \,.
\end{align}
As a final bit of notation let us define
 \begin{align}
  \accentset{1}{Q}_{\lambda \mu \nu} =  2 \frac{\partial x^{\kappa}}{\partial \xi^{\sigma}} \partial_{\lambda} \partial_{(\mu} \xi^{\sigma} g_{\nu) \kappa} 
   \end{align}
such that in an arbitrary gauge the non-metricity tensor can be written as
\begin{align}
  \label{Q_decomposition}
  Q_{\lambda \mu \nu} = \accentset{\circ}{Q}_{\lambda \mu \nu} + \accentset{1}{Q}_{\lambda \mu \nu} \,.
\end{align}
Unfortunately, we cannot avoid this somewhat cumbersome notation which will result in identifying the necessary boundary terms.

The boundary term~(\ref{equiv12}) is a linear combination of the two non-metricity pieces, moreover the partial derivative is also linear, hence one finds straightforwardly
\begin{align}
  B_Q = \frac{1}{\sqrt{-g}} \partial_{\kappa} \Big( \sqrt{-g}( \mathring{Q}_{\mu}{}^{\mu \kappa} - \mathring{Q}^{\kappa \mu}{}_{\mu}) \Big) +  \frac{1}{\sqrt{-g}} \partial_{\kappa} \Big(\sqrt{-g}
(\overset{1}{Q}{}_{\mu}{}^{\mu \kappa} -\overset{1}{Q}{}^{\kappa \mu}{}_{\mu} ) \Big)
 =: B^{(0)}_{Q} +B^{(1)}_{Q} \,,
\end{align}
which introduces two boundary terms related to the two non-metricity pieces, respectively.

The first of these boundary terms only depends on the metric and its derivatives. We find the first term to be equal to our boundary term $\ourB$, up to a minus sign
\begin{align}
B^{(0)}_{Q} &=  \frac{1}{\sqrt{-g}} \partial_{\kappa} \Big( \sqrt{-g}( \mathring{Q}_{\mu}{}^{\mu \kappa} - \mathring{Q}^{\kappa \mu}{}_{\mu}) \Big) \nonumber \\
 &=  -\frac{1}{\sqrt{-g}} \partial_{\kappa} \Big( \sqrt{-g} (g^{\mu \lambda} g^{\kappa \gamma} \partial_{\mu} g_{\lambda \gamma} - g^{\kappa \lambda} g^{\mu \gamma} \partial_{\lambda} g_{\gamma \mu} )\Big) \nonumber \\
 &= -\frac{1}{\sqrt{-g}} \partial_{\kappa} \Big( \sqrt{-g} \big( g^{\mu \lambda} \Gamma^{\kappa}_{\mu \lambda} + g^{\kappa \gamma} \Gamma^{\mu}_{\mu \gamma} - g^{\kappa \lambda} \Gamma^{\gamma}_{\gamma \lambda} - g^{\kappa \lambda} \Gamma^{\gamma}_{\gamma \lambda}  \big)\Big) \nonumber \\
 &= -\frac{1}{\sqrt{-g}} \partial_{\kappa} \Big(\sqrt{-g}\big(g^{\mu \lambda} \Gamma^{\kappa}_{\mu \lambda} - g^{\kappa \lambda} \Gamma^{\gamma}_{\gamma \lambda}\big)\Big) = -\ourB  \,,
\end{align}
where the $\Gamma$ stands for the usual Levi-Civita connection. Similarly for the other term
\begin{align}
 B^{(1)}_{Q} = \frac{1}{\sqrt{-g}} \partial_{\kappa} \big(\sqrt{-g}(\overset{1}{Q}{}_{\mu}{}^{\mu \kappa} -\overset{1}{Q}{}^{\kappa \mu}{}_{\mu})\big)
 = \frac{1}{\sqrt{-g}} \partial_{\kappa} \big(\sqrt{-g} \frac{\partial x^{\kappa}}{\partial \xi^{\sigma}} \partial^{\lambda} \partial_{\lambda} \xi^{\sigma} - \sqrt{-g} \frac{\partial x^{\rho}}{\partial \xi^{\sigma}} \partial^{\kappa} \partial_{\rho} \xi^{\sigma} \big)\,.
\end{align}
Therefore we have the identity
\begin{align}
  B_{Q} = B_{Q}^{(0)} +  B_{Q}^{(1)} = -\ourB +  B_{Q}^{(1)} \,.
\end{align}
As before, we define the difference between the non-metricity boundary term and the curvature boundary terms to be $b_{Q}$ to be consistent with the previous notation. It turns out that $b_{Q}$ happens to be $B_{Q}^{(1)}$. We write
\begin{align}
  b_{Q} := B_{Q} + \ourB = B_{Q}^{(1)} \,.
\end{align}
We are now ready to go back to~(\ref{equiv8a}) which gives
\begin{align}
  0 = \ourG + \ourB + Q + B_{Q}^{(0)} +  B_{Q}^{(1)} = \ourG + Q + b_{Q}
  \qquad \Rightarrow \qquad \ourG = - Q - b_{Q} \,.
\end{align}
This is the analogue symmetric teleparallel result of~(\ref{G_equiv}) which was derived in the teleparallel setting. Again, we note that $\ourG$ and $Q$ differ by a peculiar boundary term not previously identified. It is exactly this term which vanishes in the coincident gauge. It should be noted that the term $b_Q$ is a true Lorentz scalar but a coordinate pseudo scalar.

\subsection{Equivalence -- summary discussion}

The calculations of the preceding sections can all be summarised in the following geometrical identity which holds for globally flat affine spaces
\begin{align}
  0 = \underbrace{\ourG + \ourB}_{\mathrm{GR}} + \underbrace{T - B_T}_{\mathrm{TEGR}} + \underbrace{Q + B_{Q}}_{\mathrm{STEGR}}{}+{} \ourC\,.
\end{align}
Using our newly identified boundary terms we rewrite this as
\begin{align}
  0 = \ourG + b_G + T + b_T + Q + b_Q + \ourC\,,
\end{align}
where we introduced the notation $b_G = -\ourB$. It should now be clear that these boundary terms play a crucial role in determining the geometrical setting in which one operates. When torsion and non-metricity are assumed to vanish, one works in the standard General Relativity framework. Note that in this case the left-hand side would not be zero but equal to the Ricci scalar. Assuming that non-metricity vanishes gives TEGR, whereas the vanishing of torsion gives STEGR. It should also be noted that the cross term $\ourC$~(\ref{cross1}) vanishes when either torsion or non-metricity vanish. Therefore this term is only of importance when theories which contain both quantities are considered.

Finally one can propose the following theory 
\begin{align}
  S = \int f(\ourG,T,Q,b_G,b_T,b_Q,\ourC) \sqrt{-g}\, d^4x \,,
  \label{eq:propmodel}
\end{align}
where $f$ is a sufficiently regular function of all its variables. This setup is quite general and relates to various well-known theories as shown in Fig.~\ref{fig2}. The visualisation neglects matter couplings in the different settings, which will be discussed in the next section. We suppressed the dynamical variable in action~(\ref{eq:propmodel}) as the boundary terms now become important. Recall that the definition of the boundary term $b_T$ requires the use of tetrads, it cannot be expressed in terms of the metric alone. This means that one can always choose the tetrad to be the dynamical variable but there are model where one can work with the metric directly. 

\begin{figure}[!tbh]
\centering
\begin{tikzpicture}
  \matrix (m) [matrix of math nodes,row sep=5em,column sep=3em,minimum width=2em]
  {\mbox{} & f(\ourG,T,Q,b_G,b_T,b_Q,\ourC) & \mbox{} & \text{\begin{minipage}[r]{8em}generalised metric affine theories\end{minipage}} \\
    \mbox{} & f_{||}(\ourG,T,Q,b_G,b_T,b_Q,\ourC) & \mbox{} & \mbox{} \\
    \text{$f(T)$ \& TEGR} & \mbox{} & \text{$f(Q)$ \& STEGR} & \mbox{} \\};
  \path[-stealth]
  (m-1-2) edge node [above] {$\bar{R}_{\mu\nu\lambda\kappa}\neq 0$} (m-1-4)
  (m-1-2) edge node [left] {$\bar{R}_{\mu\nu\lambda\kappa}=0$} (m-2-2)
  (m-2-2) edge node [left] {$Q_{\mu\nu\lambda}=0\ {}$} (m-3-1)
  (m-2-2) edge node [right] {${}\ T_{\mu\nu\lambda}=0$} (m-3-3);
\end{tikzpicture}
\caption{Relationship of $f(\ourG,T,Q,b_G,b_T,b_Q,\ourC)$ gravity with other modified theories.}
\label{fig2}
\end{figure}

The mentioned generalised metric affine theories were thoroughly studied in~\cite{Hehl:1994ue}, top right in Fig.~\ref{fig2}, where the gravitational Lagrangian was assumed to be a function of the metric, the tetrads and the connection, together with their first derivatives. The gravitational Lagrangian was then assumed to be invariant under tetrad deformations which restricted it to be a function of the metric tensor and tetrad, in addition to non-metricity, torsion and curvature. Clearly, our proposed model~(\ref{eq:propmodel}) deviates from this very assumption and therefore yields distinct theories.

\section{Conclusions and discussions}
\label{sec:disc}

The starting point of this work was the Einstein or Gamma squared action which is based on the pseudo scalar $\ourG$ which differs from the Ricci scalar $R$, a true geometrical scalar, by a boundary term $\ourB$. This boundary term is also a pseudo scalar but does not contribute to the equations of motion of the theory. These pseudo scalars are not invariant under arbitrary coordinate transformations. We follow recent approaches to modified theories of gravity and introduce a theory based on the function $f(\ourG,\ourB)$. Both pseudo scalars can be expressed in terms of the metric $g_{\mu\nu}$ alone which means that the metric is the dynamical variable of the gravitational action. Consequently, in this approach matter couplings are straightforward to implement and one can follow the minimal coupling procedure without loss of generality. This is an important point as matter couplings in theories based on more general connections are well-known to become problematic when spin $1/2$ matter is taken into account. The Dirac action depends explicitly on the connection. In theories with torsion, for example, the Levi-Civita connection differs from the affine connection so that the Dirac action could be written in two different ways, leading potentially to two different theories. These subtle issues are fortunately absent in our approach.

A model based on $f(\ourG,\ourB)$ does no longer differ from a true scalar by only a boundary term, therefore one has to be particularly careful when formulating the resulting field theory. To be able to formulate a well defined variational principle for the complete theory, gravity plus matter, we must eventually assume that the total action is invariant under diffeomorphisms. By relaxing the condition that both the gravitational and matter actions must individually be invariant under diffeomorphisms, we can consider a wider class of theories than is usually studied. Similarly, the generalised conservation equations we derived are less restrictive than in other theories of modified gravity but still reduce to those of GR in the appropriate limits.

We also considered the simpler model $f(\ourG)$ gravity which shares many features with $f(Q)$ gravity and can be seen as an analogue to $f(T)$ gravity. For instance, $f(T)$ gravity is invariant under diffeomorphisms but not invariant under local Lorentz transformation. However, $f(\ourG)$ gravity is invariant under local Lorentz transformation but not invariant under diffeomorphisms. All known conceptual issue that appear in $f(T)$ gravity appear to have an analogue issue in our setup. As we have shown in Section~\ref{subsec:tegr}, we can identify a new boundary which is neither a coordinate scalar nor a Lorentz scalar. This object $b_T$, see Eq.~(\ref{def:bt}), contains information about the tetrad and the chosen coordinates, and it will be interesting to study this object further.

As a by-product of studying this simpler model, which gave cosmological field equations equivalent to those of $f(T)$ and $f(Q)$ gravity, we explored in more detail the possible differences between these three theories. A somewhat surprising result was that we could establish their equivalence despite their different geometrical interpretations and different dynamical variables. When working with suitably chosen coordinates, similar to the `good tetrads' in $f(T)$ gravity, we find field equations which appear to be covariant. In particular, these field equations \emph{imply} the usual conservation equation for the matter, again similar to the analogue situations in $f(T)$ and $f(Q)$ gravity. We also identified some new boundary terms that are neither coordinate scalars nor Lorentz scalars; these terms are sensitive to both the choice of coordinates and the choice of frames. It would be most interesting to understand the role played by these terms in different modified gravity models. Related to this issue is the so-called remnant group of $f(T)$ gravity~\cite{Chen:2014qtl,Ferraro:2014owa} which leaves the field equation unchanged. In our model one would instead deal with restricted coordinate transformations that would not affect the field equations in certain settings. This is an interesting aspect of our model that should be investigated further. 

In order to study the many different aspects of our proposed model, it appears to be sensible to consider the following framework:
\begin{align*}
  f(\ourG,\ourB) = \ourG + \mathfrak{f}(\ourG,\ourB)\,,
\end{align*}
for the gravitational part of the action and
\begin{align*}
  L_{\rm matter} = L_{\rm min\ matter}[g_{\mu\nu},\Phi] + L_{\rm non-min\ matter}[g_{\mu\nu},\Phi,\ourG,\ourB] \,,
\end{align*}
for the matter part. Isolating $\ourG$ in the gravitational action makes it straightforward to consider the limit of General Relativity. The inclusion of non-minimal coupling terms~\cite{Bertolami:2007gv} makes it possible to study some interesting models, for instance, where a scalar field couples to a boundary term~\cite{Bahamonde:2015hza}. If we were to consider couplings of the form $\alpha \phi^2 \ourB$ and $\beta \phi^2 \ourB$, where $\alpha$ and $\beta$ are coupling constants and $\phi$ is a scalar field, then the corresponding non-minimal matter action would not be a true coordinate scalar, as previously discussed. It would therefore be unnatural to assume diffeomorphism invariance for this part of the action alone, which agrees with our approach of considering models where diffeomorphism invariance only holds for the total action. Assuming the flat FLRW metric, a direct calculation shows that the cosmological field equations in the presence of a non-minimally coupled scalar field indeed agree with the field equations given in~\cite{Bahamonde:2015hza}. This should not be a surprising result at this point.

A large body of models can be explored in this way, many of which will show features distinct to previously studies theories. Note that in the cosmological setting our model yields field equations identical to those of other theories, however, this is `accidental' in the sense that it requires a specific choice of tetrads and coordinates. In general one can expected different features.

The approach outlined in this work allows the possibility to study gravitational theories which break diffeomorphism invariance at certain scales, for instance very small scales where classical physics breaks down, or very large, cosmological scales. We carefully set up the required framework to study models of this type in a variety of situations. The inclusion of non-minimal matter couplings should make this framework sufficiently general for most realistic applications. The identification of various new boundary terms, not previously discussed in the literature, allowed us to link our model to many previously studied modified gravity theories.

\subsection*{Acknowledgements}

We are grateful to the organisers and participants of the Teleparallel Gravity Workshop 2020, Tartu, Estonia for the interesting discussions which motivated parts of this work. The authors would like to thank Sebastian Bahamonde, Alan Coley, Rafael Ferraro, Franco Fiorini and Manos Saridakis for their helpful comments. Erik Jensko is supported by EPSRC Doctoral Training Programme (EP/R513143/1).

\appendix

\section{Variations of the Einstein action}
\label{Appendix_Einstein}

\subsection{The Einstein action}

Here we calculate the variations of the Einstein action where we naturally introduce the objects $M^{\mu \nu}{}_{\lambda}$ and $E^{\mu \nu \lambda}$. The metric variation of the Einstein action~(\ref{Einstein_Action}) and the matter action $S_{\rm{matter}}[g_{\mu \nu},\Phi]$ can be written as
\begin{align}
  \delta S &= \delta S_{\rm{E}}[g_{\mu \nu}] + \delta S_{\rm{matter}}[g_{\mu \nu},\Phi] = 0
  \nonumber \\ &=
  \frac{1}{2 \kappa} \int d^4x \Big[ \delta \sqrt{-g} \ourG + \sqrt{-g} \delta \ourG +
    2 \kappa \frac{\delta S_{\rm{matter}}}{\delta g^{\mu \nu}}\delta g^{\mu \nu} \Big] \,,
  \label{Einstein_Action_Matter_Variation}
\end{align}
with $\ourG$ previously introduced in~(\ref{G_bulk}). Focusing on the gravitational action and dropping the factor of $2 \kappa$ we have
\begin{align}
  \label{Variation_of_Einstein_action}
  \delta S_{\rm{E}} = \int \delta \sqrt{-g} g^{\mu \nu} (\Gamma_{\mu \nu})^2\, d^4x +
  \int \sqrt{-g} \ \delta g^{\mu \nu} (\Gamma_{\mu \nu})^2\, d^4x +
  \int \sqrt{-g} g^{\mu \nu} \delta (\Gamma_{\mu \nu})^2\, d^4x \,,
\end{align}
where here $(\Gamma_{\mu \nu})^2$ is short for $\Gamma^{\sigma}_{\lambda \mu} \Gamma^{\lambda}_{\nu \sigma} - \Gamma^{\sigma}_{\mu \nu} \Gamma^{\lambda}_{\sigma \lambda}$. The first two terms are
\begin{align}
  \label{metric_determinant_variation}
  \int \delta \sqrt{-g} g^{\mu \nu} (\Gamma_{\mu \nu})^2 d^4x &= -\frac{1}{2} \int \sqrt{-g} \ \delta g^{\rho \sigma} g_{\rho \sigma} g^{\mu \nu} (\Gamma^{\gamma}_{\lambda \mu} \Gamma^{\lambda}_{\gamma \nu} - \Gamma^{\gamma}_{\mu \nu} \Gamma^{\lambda}_{\gamma \lambda} ) d^4 x \,, \\
  \label{metric_tensor_variation}
  \int \sqrt{-g} \ \delta g^{\mu \nu} (\Gamma_{\mu \nu})^2 d^4x &= \int \sqrt{-g} \ \delta g^{\rho \sigma} (\Gamma^{\gamma}_{\lambda \rho} \Gamma^{\lambda}_{\gamma \sigma} - \Gamma^{\gamma}_{\rho \sigma} \Gamma^{\lambda}_{\gamma \lambda} ) d^4 x \,.
\end{align}
Next we expand $g^{\mu \nu} \delta (\Gamma_{\mu \nu})^2$ to arrive at
\begin{align}
    g^{\mu \nu} \delta (\Gamma_{\mu \nu})^2 &=g^{\mu \nu} \big( \delta \Gamma^{\sigma}_{\lambda \mu} \Gamma^{\lambda}_{\sigma \nu} + \delta \Gamma^{\lambda}_{\sigma \nu} \Gamma^{\sigma}_{\lambda \mu} - \delta \Gamma^{\sigma}_{\mu \nu} \Gamma^{\lambda}_{\sigma \lambda} - \delta \Gamma^{\lambda}_{\sigma \lambda} \Gamma^{\sigma}_{\mu \nu} \big) \nonumber \\
    &= g^{\mu \nu} (\delta^{\sigma}_{\mu} \Gamma^{\rho}_{\kappa \nu} + \delta^{\sigma}_{\nu} \Gamma^{\rho}_{\kappa \mu} - \delta^{\rho}_{\mu} \delta^{\sigma}_{\nu} \Gamma^{\lambda}_{\kappa \lambda} - \delta^{\sigma}_{\kappa} \Gamma^{\rho}_{\mu \nu}) \delta \Gamma^{\kappa}_{\rho \sigma} \nonumber \\
    &= M^{\rho \sigma}{}_{\kappa} \delta \Gamma^{\kappa}_{\rho \sigma} \,.
\end{align}
Here we introduced the object 
\begin{align}
  \label{M_Appendix}
 M^{\alpha \beta}{}_{\gamma} :=  2 g^{\nu (\beta} \Gamma^{\alpha)}_{\gamma \nu} - g^{\alpha \beta} \Gamma^{\lambda}_{\gamma \lambda} - g^{\mu \nu}  \delta^{(\beta}_{\gamma} \Gamma^{\alpha)}_{\mu \nu} \,,
\end{align}
which is just the variation of $\ourG$ with respect to the connection
\begin{align}
    \frac{\partial \mathbf{G} }{\partial \Gamma^{\gamma}_{\alpha \beta}} = M^{\alpha \beta}{}_{\gamma} \,.
\end{align}
The object $M^{\alpha \beta}{}_{\gamma}$ is constructed to be symmetric over its first two indices to match the symmetry of the connection.

Expanding the variation of the Christoffel connection in terms of metric variations gives
\begin{align}
  \delta \Gamma^{\kappa }_{\rho \sigma } &= \frac{1}{2} \delta g^{\kappa \lambda } (g_{\rho \lambda ,\sigma } + g_{\sigma \lambda ,\rho } - g_{\rho \sigma ,\lambda } ) + \frac{1}{2} g^{\kappa \lambda } (\delta g_{\rho \lambda ,\sigma } + \delta g_{\sigma \lambda ,\rho } - \delta g_{\rho \sigma ,\lambda } )
  \nonumber \\
  &= \delta g^{\kappa \lambda } g_{\zeta \lambda } \Gamma^{\zeta }_{\rho \sigma } + \frac{1}{2} g^{\kappa \lambda } (\delta g_{\rho \lambda ,\sigma } + \delta g_{\sigma \lambda ,\rho } - \delta g_{\rho \sigma ,\lambda })
  \nonumber \\
  &= \delta g^{\kappa \lambda } g_{\zeta \lambda } \Gamma^{\zeta }_{\rho \sigma } + \frac{1}{2} g^{\kappa \lambda } (\delta g_{\alpha \beta , \gamma} \Delta^{\alpha \beta \gamma}_{\rho  \sigma  \lambda }) \,,
  \label{Chr_variation}
\end{align}
where we also define
\begin{align}
  \Delta^{\alpha \beta \gamma}_{\rho \sigma \lambda} =
  \delta^{\alpha}_{\{\lambda} \delta^{\beta}_{\rho} \delta^{\gamma}_{\sigma \}} =
  \delta^{\alpha}_{\lambda} \delta^{\beta}_{\rho} \delta^{\gamma}_{\sigma} +
  \delta^{\alpha}_{\sigma} \delta^{\beta}_{\lambda} \delta^{\gamma}_{\rho} -
  \delta^{\alpha}_{\rho} \delta^{\beta}_{\sigma} \delta^{\gamma}_{\lambda} \,,
\end{align}
which simply permutes indices using the \emph{Schouten bracket}, see~\cite{JS1954}. Again we note that only the symmetric part over the indices $\alpha \, \beta$ of $\Delta^{\alpha \beta \gamma}_{\rho \sigma \lambda}$ in~(\ref{Chr_variation}) contributes.

Putting the above equations back into the integrand, we can write the last term of~(\ref{Variation_of_Einstein_action}) as
\begin{align}
  \int \sqrt{-g} g^{\mu \nu} \delta (\Gamma_{\mu \nu})^2\, d^4x &=
  \int \sqrt{-g}  M^{\rho \sigma}{}_{\kappa} \delta \Gamma^{\kappa}_{\rho \sigma}\, d^4 x
  \nonumber \\ &=
  \int \sqrt{-g}  M^{\rho \sigma}{}_{\kappa} \big[  \delta g^{\kappa \lambda } g_{\zeta \lambda }
    \Gamma^{\zeta }_{\rho \sigma } + \frac{1}{2} g^{\kappa \lambda } (\delta g_{\alpha \beta , \gamma}
    \Delta^{\alpha \beta \gamma}_{\rho  \sigma  \lambda })\big]\, d^4 x \,.
\end{align}
To simplify the second term, we define the new object $E$ as the permutation of $M^{\rho \sigma \lambda}$ given by
\begin{align}
  E^{\alpha \beta \gamma} := \Delta^{\alpha \beta \gamma}_{\rho \sigma \lambda} M^{\rho \sigma \lambda} &= M^{\beta \gamma \alpha} + M^{\gamma \alpha \beta} - M^{\alpha \beta \gamma}
  \nonumber \\ &=
  2 g^{\nu \alpha} g^{\beta \mu} \Gamma^{\gamma}_{\mu \nu} -
  2 g^{\gamma (\alpha} g^{\beta) \mu} \Gamma^{\lambda}_{\lambda \mu} +
  g^{\alpha \beta} g^{\mu \gamma} \Gamma^{\lambda}_{\lambda \mu} -
  g^{\alpha \beta} g^{\mu \nu} \Gamma^{\gamma}_{\mu \nu} \,,
\end{align}
allowing us to write
\begin{align}
  \label{Gammapre}
  \int \sqrt{-g} g^{\mu \nu} \delta (\Gamma_{\mu \nu})^2 d^4x = \int \sqrt{-g} \Big(\delta g^{\kappa \lambda } g_{\zeta \lambda } M^{\rho \sigma}{}_{\kappa} \Gamma^{\zeta }_{\rho \sigma }+ \frac{1}{2} E^{\alpha \beta \gamma} \partial_{\gamma}\delta g_{\alpha\beta} \Big)\, d^4 x \,.
\end{align}
Integrating the second term by parts and dropping the boundary terms, then rewriting the metric variation in terms of the inverse metric gives
\begin{align}
  \label{Gamma_squared_variation}
  \int \sqrt{-g} g^{\mu \nu} \delta (\Gamma_{\mu \nu})^2\, d^4x =
  \int \delta g^{\rho \sigma} \Big[
    \sqrt{-g}  g_{\zeta \sigma}  M^{\mu \nu}{}_{\rho}
    \Gamma^{\zeta}_{\mu \nu} + \frac{1}{2} g_{\alpha \rho} g_{\beta \sigma}
    \partial_{\gamma} ( \sqrt{-g} E^{\alpha \beta \gamma}) \Big] \, d^4 x \,.
\end{align}
As we wish to obtain the Einstein tensor, we'll simply expand everything in terms of the connection and its derivatives.
The first term in the integrand of~(\ref{Gamma_squared_variation}) can be expanded as
\begin{align}
        \delta g^{\rho \sigma} \sqrt{-g}\, g_{\zeta  \sigma}  M^{\mu \nu}{}_{\rho}
    \Gamma^{\zeta }_{\mu \nu}  =  \delta g^{\rho \sigma} \sqrt{-g} \Big( 2 g^{\nu  \eta} g_{\sigma \zeta } \Gamma^{\zeta }_{\mu  \nu } \Gamma^{\mu }_{\rho \eta} -  g^{\mu \nu } g_{\sigma \zeta } \Gamma^{\zeta }_{\mu \nu } \Gamma^{\lambda}_{\lambda \rho} - g^{\gamma \eta} g_{\sigma \zeta } \Gamma^{\zeta }_{\mu  \rho} \Gamma^{\mu }_{\gamma \eta} \Big) \,.
\end{align}
The second term of~(\ref{Gamma_squared_variation}) involves taking the partial derivative of $E^{\alpha \beta \gamma}$, which leads to the following expression
\begin{multline} \label{DerivativeE_final}
    \frac{1}{2} \delta g^{\rho \sigma}  g_{\alpha \rho} g_{\beta \sigma} \partial_{\gamma} ( \sqrt{-g} E^{\alpha \beta \gamma}) =  \sqrt{-g} \delta g^{\rho \sigma} \Big( 
      2\Gamma^{\lambda}_{\kappa \lambda} \Gamma^{\kappa }_{\rho \sigma}
     -2\Gamma^{\gamma}_{\sigma \nu} \Gamma^{\nu}_{\gamma \rho}
     -g_{\sigma \rho} g^{\mu \nu} \Gamma^{\eta}_{\gamma \eta} \Gamma^{\gamma}_{\mu \nu}
     -2 g_{\alpha \rho} g^{\epsilon \nu} \Gamma^{\gamma}_{\sigma \nu} \Gamma^{\alpha}_{\gamma \epsilon} \\
     +  g_{\alpha \rho} g^{\eta \gamma} \Gamma^{\alpha}_{\gamma \eta} \Gamma^{\lambda}_{\sigma \lambda} 
     +g_{\alpha \rho} g^{\mu \nu} \Gamma^{\gamma}_{\mu \nu} \Gamma^{\alpha}_{\gamma \sigma} 
     + g_{\rho \sigma} g^{\eta \nu} \Gamma^{\gamma}_{\mu \nu} \Gamma^{\mu}_{\gamma \eta} \\
     + \partial_{\gamma} \Gamma^{\gamma}_{\rho \sigma} -  \partial_{\rho} \Gamma^{\lambda}_{\sigma \lambda} 
    + \frac{1}{2} g_{\sigma \rho} g^{\kappa  \gamma} \partial_{\gamma} \Gamma^{\lambda}_{\kappa  \lambda} - \frac{1}{2} g_{\rho \sigma} g^{\mu \nu} \partial_{\gamma} \Gamma^{\gamma}_{\mu \nu}
     \Big) \,.
\end{multline}
The full $(\Gamma_{\mu \nu})^2$ variation becomes
\begin{multline}
  \label{Final_gamma_squared_variation}
  \int  \sqrt{-g} g^{\mu \nu} \delta (\Gamma_{\mu \nu})^2\, d^4x =
  \int \delta g^{\rho \sigma} \sqrt{-g} \Big( 2 \Gamma^{\eta}_{\gamma \eta} \Gamma^{\gamma}_{\rho \sigma} 
    -2  \Gamma^{\gamma}_{\sigma \nu} \Gamma^{\nu}_{\gamma \rho}
    + g_{\rho \sigma} g^{\eta \nu} \Gamma^{\gamma}_{\mu \nu} \Gamma^{\mu}_{\gamma \eta} 
    - g_{\sigma \rho} g^{\mu \nu} \Gamma^{\eta}_{\gamma \eta} \Gamma^{\gamma}_{\mu \nu}  \\
    + \partial_{\gamma} \Gamma^{\gamma}_{\rho \sigma}
    -  \partial_{\rho} \Gamma^{\lambda}_{\sigma \lambda} 
    + \frac{1}{2} g_{\sigma \rho} g^{\kappa \gamma} \partial_{\gamma} \Gamma^{\lambda}_{\kappa \lambda} - \frac{1}{2} g_{\rho \sigma} g^{\mu \nu} \partial_{\gamma} \Gamma^{\gamma}_{\mu \nu} \Big)\, d^4 x \,.
\end{multline}

\subsection{Recovering the Einstein field equations}
\label{Appendix_EFE}

Putting together equations~(\ref{metric_determinant_variation}),~(\ref{metric_tensor_variation}) and~(\ref{Final_gamma_squared_variation}), we get a final expression for the variation of the Einstein action
\begin{multline} 
    \delta S_{\rm{E}} = \int \sqrt{-g} \delta g^{\rho \sigma} \Big[\Gamma^{\eta}_{\gamma \eta} \Gamma^{\gamma}_{\rho \sigma} -  \Gamma^{\gamma}_{\sigma \nu} \Gamma^{\nu}_{\gamma \rho} +
    \frac{1}{2} g_{\rho \sigma} g^{\eta \nu} \Gamma^{\gamma}_{\mu \nu} \Gamma^{\mu}_{\gamma \eta} -
    \frac{1}{2} g_{\sigma \rho} g^{\mu \nu} \Gamma^{\eta}_{\gamma \eta} \Gamma^{\gamma}_{\mu \nu}  \\
     + \partial_{\gamma} \Gamma^{\gamma}_{\rho \sigma}
    -  \partial_{\rho} \Gamma^{\lambda}_{\sigma \lambda} 
    + \frac{1}{2} g_{\sigma \rho} g^{\kappa \gamma} \partial_{\gamma} \Gamma^{\lambda}_{\kappa \lambda} - \frac{1}{2} g_{\rho \sigma} g^{\mu \nu} \partial_{\gamma} \Gamma^{\gamma}_{\mu \nu}
    \Big]\, d^4x \,.
\end{multline}
The term in the square brackets is symmetric over $\rho \, \sigma$, we left out the implicit symmetry brackets to make the calculation clearer. We then recognise this to be the Einstein tensor,
\begin{align}
    \delta S_{\rm{E}} = \int \sqrt{-g} \delta g^{\rho \sigma} \big(R_{\rho \sigma} - \frac{1}{2} g_{\rho \sigma} R \big)\, d^4 x  = \int \sqrt{-g} \delta g^{\rho \sigma} G_{\rho \sigma}\, d^4 x \,.
\end{align}
The full action variation~(\ref{Einstein_Action_Matter_Variation}) then leads to
\begin{align}
  \delta S = \delta S_{\rm{E}}[g_{\mu \nu}] + \delta S_{\rm{matter}}[g_{\mu \nu},\Phi] &= 0
  \nonumber \\
  \int \frac{1}{2 \kappa}  \sqrt{-g}  \delta g^{\rho \sigma}
  \Big( G_{\rho \sigma} + \frac{2 \kappa}{\sqrt{-g}}
  \frac{\delta L_{\rm{M}}}{\delta g^{\rho \sigma}}\Big)\, d^4x &= 0 \\
   \implies\qquad G_{\rho \sigma} =\kappa T_{\rho \sigma}& \,,
\end{align}
where the metric energy-momentum tensor is defined by
\begin{align}
  T_{\mu \nu} = -\frac{2}{\sqrt{-g}}
  \frac{\delta L_{\rm{M}}[g_{\mu \nu}, \Phi]}{\delta g^{\mu \nu}} \,.
\end{align}

\section{Variations of the \texorpdfstring{$f(\ourG,\ourB)$}{f(G,B)} action}
\label{Appendix_f(G,B)_action}

\subsection{\texorpdfstring{$f(\ourG,\ourB)$}{f(G,B)} action}

Here we calculate the variation of the $f(\ourG,\ourB)$ action~(\ref{f(G,B)_action}). We begin with equations~(\ref{f(G,B)_action_variation}) and~(\ref{f(G,B)_variation}), ignoring for now the factor of $2 \kappa$
\begin{align}
  \delta S_{\rm grav} = \int \left[\delta f(\ourG,\ourB) \sqrt{-g} -
  \frac{1}{2}  f(\ourG,\ourB)\sqrt{-g} g_{\rho \sigma} \delta g^{\rho \sigma} \right] d^4 x \,,
  \nonumber \\
  \delta  f(\ourG,\ourB) = \frac{\partial  f(\ourG,\ourB)}{\partial \ourG} \delta \ourG +
  \frac{\partial  f(\ourG,\ourB)}{\partial \ourB} \delta \ourB \,,
  \label{Field_equation_derivation_Appendix_eq1}
\end{align}
where $\ourG$ and $\ourB$ are defined in~(\ref{G_bulk}) and~(\ref{B_boundary})
\begin{align*}
  \ourG = g^{\mu \nu}\Big( \Gamma^{\lambda}_{\mu \sigma} \Gamma^{\sigma}_{\lambda \nu} - \Gamma^{\sigma}_{\mu \nu} \Gamma^{\lambda}_{\lambda \sigma} \Big)\,, \qquad 
  \ourB = \frac{1}{\sqrt{-g}} \partial_{\nu}
  \Big( \frac{\partial_{\mu}(g g^{\mu \nu})}{\sqrt{-g}} \Big) \,.
\end{align*}
We will use the following shorthand notation for derivatives, $f_{,\ourG} = \partial f(\ourG,\ourB)/\partial \ourG$, $f_{,\ourB} = \partial f(\ourG,\ourB)/\partial \ourB$, etc. First we will look at the variation of the bulk term and then the variation of the boundary term. Note that most of the calculations for $\delta \ourG$ were covered in the previous appendix.

\subsection{Bulk term}
\label{Appendix_Bulk_term}

In Appendix~\ref{Appendix_Einstein} we calculated the variation of the bulk term, so using equations~(\ref{metric_tensor_variation}) and~(\ref{Gammapre}) we can write
\begin{align}
    \delta \ourG = \delta g^{\rho \sigma} \Big(\Gamma^{\gamma}_{\lambda \rho} \Gamma^{\lambda}_{\gamma \sigma} - \Gamma^{\gamma}_{\rho \sigma} \Gamma^{\lambda}_{\gamma \lambda} + g_{\gamma \sigma}
    \Gamma^{\gamma}_{\mu \nu} M^{\mu \nu }{}_{\rho} \Big) + \frac{1}{2} E^{\alpha \beta \gamma} \partial_{\gamma} \delta g_{\alpha \beta} \,.
\end{align}
The first term of the integrand~(\ref{Field_equation_derivation_Appendix_eq1}) is then
\begin{align}
  \label{deltaGintegral_Appendix}
  \sqrt{-g} f_{,\ourG} \delta \ourG = \sqrt{-g} f_{,\ourG}
  \Big[ \delta g^{\rho \sigma} \Big(\Gamma^{\gamma}_{\lambda \rho} \Gamma^{\lambda}_{\gamma \sigma} -
    \Gamma^{\gamma}_{\rho \sigma} \Gamma^{\lambda}_{\gamma \lambda} + g_{\gamma \sigma}
    \Gamma^{\gamma}_{\mu \nu} M^{\mu \nu}{}_{\rho} \Big) +
    \frac{1}{2} E^{\alpha \beta \gamma} \partial_{\gamma} \delta g_{\alpha \beta} \Big] \,.
\end{align}
Using the definition of $ M^{\mu \nu}{}_{\rho} $~(\ref{M_Appendix}), we can write the term in the curved brackets above as
\begin{align} \label{f_G_1_Appendix}
    \sqrt{-g}  f_{, \ourG} \delta g^{\rho \sigma} \Bigg[ 
     (\Gamma^{\gamma}_{\lambda \rho} \Gamma^{\lambda}_{\gamma \sigma} - \Gamma^{\gamma}_{\rho \sigma} \Gamma^{\lambda}_{\gamma \lambda} ) +
     \Big( 2 g^{\nu \lambda} g_{\sigma \gamma} \Gamma^{\gamma}_{\mu \nu} \Gamma^{\mu}_{\rho \lambda} -  g^{\mu \nu} g_{\sigma \gamma} \Gamma^{\gamma}_{\mu \nu} \Gamma^{\lambda}_{\lambda \rho} - g^{\gamma \lambda} g_{\sigma \nu} \Gamma^{\nu}_{\mu \rho} \Gamma^{\mu}_{\gamma \lambda} \Big) \Bigg] \,.
\end{align}
Finally, the last term in~(\ref{deltaGintegral_Appendix}) is integrated by parts
\begin{multline}
  \frac{1}{2} \int \sqrt{-g} f_{,\ourG} E^{\alpha \beta \gamma} (\delta g_{\alpha \beta , \gamma})\, d^4x =
  \textrm{boundary terms} - \frac{1}{2} \int \delta g_{\alpha \beta} \partial_{\gamma}
  \big( \sqrt{-g} f_{,\ourG} E^{\alpha \beta \gamma} \big)\, d^4x \\ =
  \frac{1}{2} \int \Big[ \sqrt{-g} \delta g^{\rho \sigma} \partial_{\gamma}(f_{,\ourG})
    E_{\rho \sigma}{}^{\gamma} + \delta g^{\alpha \beta} g_{\alpha \rho} g_{\beta \sigma}
    f_{, \ourG} \partial_{\gamma}(\sqrt{-g} E^{\alpha \beta \gamma}) \Big]\, d^4 x \,,
  \label{f_G_E_Appendix}
\end{multline}
where the boundary term proportional to $\delta g_{\alpha \beta}$ does not contribute. We already calculated $\partial_{\gamma}(\sqrt{-g} E^{\alpha \beta \gamma})$, so the second term in the final line of~(\ref{f_G_E_Appendix}) is just $f_{, \ourG}$ times~(\ref{DerivativeE_final})
\begin{multline}
  \label{Derivative_E}
    \frac{1}{2} \delta g^{\rho \sigma}  g_{\alpha \rho} g_{\beta \sigma} \partial_{\gamma} f_{, \ourG} ( \sqrt{-g} E^{\alpha \beta \gamma}) =  \sqrt{-g} \delta g^{\rho \sigma} f_{, \ourG} \Bigg[ 
      2\Gamma^{\lambda}_{\kappa \lambda} \Gamma^{\kappa}_{\rho \sigma}
     -2\Gamma^{\gamma}_{\sigma \nu} \Gamma^{\nu}_{\gamma \rho} \\ 
     -g_{\sigma \rho} g^{\mu \nu} \Gamma^{\eta}_{\gamma \eta} \Gamma^{\gamma}_{\mu \nu}
     -2 g_{\alpha \rho} g^{\epsilon \nu} \Gamma^{\gamma}_{\sigma \nu} \Gamma^{\alpha}_{\gamma \epsilon} 
     +  g_{\alpha \rho} g^{\eta \gamma} \Gamma^{\alpha}_{\gamma \eta} \Gamma^{\lambda}_{\sigma \lambda} 
     +g_{\alpha \rho} g^{\mu \nu} \Gamma^{\gamma}_{\mu \nu} \Gamma^{\alpha}_{\gamma \sigma} 
     + g_{\rho \sigma} g^{\eta \nu} \Gamma^{\gamma}_{\mu \nu} \Gamma^{\mu}_{\gamma \eta} \\
     + \partial_{\gamma} \Gamma^{\gamma}_{\rho \sigma} -  \partial_{\rho} \Gamma^{\lambda}_{\sigma \lambda} 
    + \frac{1}{2} g_{\sigma \rho} g^{\kappa \gamma} \partial_{\gamma} \Gamma^{\lambda}_{\kappa \lambda} - \frac{1}{2} g_{\rho \sigma} g^{\mu \nu} \partial_{\gamma} \Gamma^{\gamma}_{\mu \nu}
     \Bigg] \,.
\end{multline}
From comparison with Appendix~\ref{Appendix_EFE}, we can immediately notice that the terms inside the square brackets of equations~(\ref{f_G_1_Appendix}) and~(\ref{Derivative_E}) almost give the Einstein tensor, but they are missing the variation of the metric determinant multiplied by $\ourG$~(\ref{metric_determinant_variation}). We therefore have that our final variation of the bulk term is given by,
\begin{align}
  \label{bulk_term_variation}
  \int \sqrt{-g} f_{, \ourG} \delta \ourG \, d^4x  = \int \sqrt{-g} \delta g^{\rho \sigma} \Big[ f_{, \ourG} \Big( G_{\rho \sigma} + \frac{1}{2} g_{\rho \sigma} \ourG \big) + \frac{1}{2} E_{\rho \sigma}{}^{\gamma} \partial_{\gamma} f_{, \ourG} \Big]\, d^4x \,.
\end{align}

\subsection{Boundary term}
\label{Appendix_Boundary_term}

The variation of the boundary term is
\begin{align}  
    \delta \ourB = \delta \Big( \frac{1}{\sqrt{-g}}\Big) \partial_{\nu} \Big( \frac{\partial_{\mu}(g g^{\mu \nu})}{\sqrt{-g}} \Big) + \frac{1}{\sqrt{-g}} \delta\Bigg( \partial_{\nu} \Big( \frac{\partial_{\mu}(g g^{\mu \nu})}{\sqrt{-g}} \Big) \Bigg) \,.
\end{align}
The first term gives
\begin{align}
    \frac{1}{2 \sqrt{-g}} g_{\alpha \beta} \delta g^{\alpha \beta}
    \partial_{\nu} \Big( \frac{\partial_{\mu}(g g^{\mu \nu})}{\sqrt{-g}} \Big) = \frac{1}{2} \delta g^{\alpha \beta} g_{\alpha \beta} \ourB \,.
\end{align}
The second term we split into three calculations
\begin{align} 
  \frac{1}{\sqrt{-g}} \delta\Bigg( \partial_{\nu} \Big( \frac{\partial_{\mu}(g g^{\mu \nu})}{\sqrt{-g}} \Big) \Bigg) = \frac{1}{\sqrt{-g}} \partial_{\nu} \Big\{
    \overbrace{\frac{\partial_{\mu} (\delta g g^{\mu \nu}) }{\sqrt{-g}}}^{1}
    + \overbrace{\frac{\partial_{\mu}(g \delta g^{\mu \nu}) }{\sqrt{-g}}}^{2}
    + \overbrace{\partial_{\mu}(g g^{\mu \nu}) \delta \Big(\frac{1}{\sqrt{-g}} \Big)}^{3} \Big\} \,,
\end{align}
where the first term is given by
 \begin{multline}
    \frac{1}{\sqrt{-g}} \partial_{\nu} \Big[   \overbrace{\frac{\partial_{\mu} (\delta g g^{\mu \nu}) }{\sqrt{-g}}}^{1} \Big] = \frac{1}{\sqrt{-g}} \partial_{\nu} 
    \Bigg\{\sqrt{-g} \delta g^{\alpha \beta} \Big[2 \Gamma^{\eta}_{\mu \eta} g_{\alpha \beta} g^{\mu \nu} \\
    + 2  \Gamma^{\eta}_{\mu \alpha} g_{\eta \beta}g^{\mu \nu} 
    -  g_{\alpha \beta} \big( \Gamma^{\mu}_{\mu \eta} g^{\eta \nu} + \Gamma^{\nu}_{\mu \eta} g^{\mu \eta} \big) \Big] + \sqrt{-g} g_{\alpha \beta} g^{\mu \nu} \partial_{\mu}\big( \delta g^{\alpha \beta} \big) \Bigg\} \,.
\end{multline}
For the second and third term we find
 \begin{align}
    \frac{1}{\sqrt{-g}} \partial_{\nu} \Big[\overbrace{\frac{\partial_{\mu}(g \delta g^{\mu \nu}) }{\sqrt{-g}} }^{2}\Big] &=  \frac{-1}{\sqrt{-g}} \partial_{\nu} \Bigg[ 
    \sqrt{-g} \Big[2 \delta g ^{\mu \nu} \Gamma^{\alpha}_{\mu \alpha} + \partial_{\mu}(\delta g^{\mu \nu}) \Big] \Bigg] \,, \\
    \frac{1}{\sqrt{-g}} \partial_{\nu} \bigg[\overbrace{\partial_{\mu}(g g^{\mu \nu}) \delta \Big(\frac{1}{\sqrt{-g}} \Big)}^{3}  \bigg] &= \frac{1}{\sqrt{-g}} \partial_{\nu} \Bigg[ \sqrt{-g} \delta g^{\alpha \beta} \Big[
    \frac{g_{\alpha \beta}}{2} \big( \Gamma^{\mu}_{\mu \eta} g^{\eta \nu} + \Gamma^{\nu}_{\mu \eta} g^{\mu \eta} \big)  - \Gamma^{\eta}_{\mu \eta} g_{\alpha \beta} g^{\mu \nu}\Big] \Bigg] \,. 
\end{align}
Collecting these together gives
\begin{multline}
    \frac{1}{\sqrt{-g}} \partial_{\nu} \Bigg\{\sqrt{-g} \bigg[ \delta g^{\alpha \beta} 
    \Big( \frac{1}{2} g_{\alpha \beta} g^{\mu \nu} \Gamma^{\eta}_{\mu \eta} - \frac{1}{2} g_{\alpha \beta} g^{\mu \eta} \Gamma^{\nu}_{\mu \eta}  + 2 g_{\eta \beta} g^{\mu \nu} \Gamma^{\eta}_{\mu \alpha} 
    - 2 \delta_{\alpha}^{\nu} \Gamma^{\gamma}_{\beta \gamma} 
    \Big) \\
    +  g_{\alpha \beta} g^{\mu \nu} \partial_{\mu}(\delta g^{\alpha \beta}) - \partial_{\mu}(\delta g^{\mu \nu}) \bigg]
    \Bigg\} \,.
\end{multline}
The total variation of $\ourB$ can then be written as
\begin{multline}
  \label{Variation_B_Appendix}
    \delta \ourB =  \frac{1}{2} \delta g^{\alpha \beta} g_{\alpha \beta} \ourB  + \frac{1}{\sqrt{-g}} \partial_{\nu} \Bigg\{ \sqrt{-g} \bigg[ \delta g^{\alpha \beta} 
    \Big( \frac{1}{2} g_{\alpha \beta} g^{\mu \nu} \Gamma^{\eta}_{\mu \eta} - \frac{1}{2} g_{\alpha \beta} g^{\mu \eta} \Gamma^{\nu}_{\mu \eta}  +  \\ 2 g_{\eta \beta} g^{\mu \nu} \Gamma^{\eta}_{\mu \alpha} 
    - 2 \delta_{\alpha}^{\nu} \Gamma^{\gamma}_{\beta \gamma} 
    \Big) 
    +  g_{\alpha \beta} g^{\mu \nu} \partial_{\mu}(\delta g^{\alpha \beta}) - \partial_{\mu}(\delta g^{\mu \nu}) \bigg]
    \Bigg\} \,.
\end{multline}
Substituting $\delta \ourB$~(\ref{Variation_B_Appendix}) into the variational integral~(\ref{Field_equation_derivation_Appendix_eq1}) yields
\begin{multline}
  \label{variation_of_B_Int_by_Parts}
    \int \sqrt{-g} f_{, \ourB} \delta \ourB \, d^4 x= \int \frac{1}{2} \sqrt{-g} f_{, \ourB}   \delta g^{\alpha \beta} g_{\alpha \beta} \ourB  \, d^4x  \\ 
  + f_{, \ourB} \partial_{\nu} \Bigg\{\sqrt{-g} \bigg[ \overbrace{\delta g^{\alpha \beta} 
    \Big( \frac{1}{2} g_{\alpha \beta} g^{\mu \nu} \Gamma^{\eta}_{\mu \eta} - \frac{1}{2} g_{\alpha \beta} g^{\mu \eta} \Gamma^{\nu}_{\mu \eta}  +  2 g_{\eta \beta} g^{\mu \nu} \Gamma^{\eta}_{\mu \alpha} 
    - 2 \delta_{\alpha}^{\nu} \Gamma^{\gamma}_{\beta \gamma} 
    \Big) }^{A} \\
    +  \overbrace{g_{\alpha \beta} g^{\mu \nu} \partial_{\mu}(\delta g^{\alpha \beta}) - \partial_{\mu}(\delta g^{\mu \nu}) }^{B} \bigg]
    \Bigg\}\, d^4x \,,
\end{multline}
where we perform integration by parts once on the $A$ terms and twice on the $B$ terms
\begin{multline}
  \label{Integration_by_parts_B_first_term_Appendix}
    \int f_{, \ourB}  \partial_{\nu} \Bigg[ \sqrt{-g} \delta g^{\alpha \beta} 
    \Big( \overbrace{\frac{1}{2} g_{\alpha \beta} g^{\mu \nu} \Gamma^{\eta}_{\mu \eta} - \frac{1}{2} g_{\alpha \beta} g^{\mu \eta} \Gamma^{\nu}_{\mu \eta}  +  2 g_{\eta \beta} g^{\mu \nu} \Gamma^{\eta}_{\mu \alpha} 
    - 2 \delta_{\alpha}^{\nu} \Gamma^{\gamma}_{\beta \gamma} 
    \Big) }^{A} \Bigg]\, d^4x = 
    \\ \textrm{boundary terms} - \int \delta g^{\alpha \beta} \partial_{\nu}(f_{, \ourB}) \Big( \frac{1}{2} g_{\alpha \beta} g^{\mu \nu} \Gamma^{\eta}_{\mu \eta} - \frac{1}{2} g_{\alpha \beta} g^{\mu \eta} \Gamma^{\nu}_{\mu \eta}  +  2 g_{\eta \beta} g^{\mu \nu} \Gamma^{\eta}_{\mu \alpha} 
    - 2 \delta_{\alpha}^{\nu} \Gamma^{\gamma}_{\beta \gamma} 
    \Big)\, d^4 x \,,
\end{multline}
\begin{multline}
  \label{integration_by_parts_B_Appendix}
    \int f_{, \ourB}  \partial_{\nu} \Bigg[ \sqrt{-g} \Big( \overbrace{g_{\alpha \beta} g^{\mu \nu} \partial_{\mu}(\delta g^{\alpha \beta}) - \partial_{\mu}(\delta g^{\mu \nu}) }^{B} \Big) \Bigg]\, d^4 x = \\
    \textrm{boundary terms} - \int \partial_{\nu} (f_{, \ourB}) \sqrt{-g} \Big( g_{\alpha \beta} g^{\mu \nu} \partial_{\mu}(\delta g^{\alpha \beta}) - \partial_{\mu}(\delta g^{\mu \nu}) \Big)\, d^4 x = \\
    \textrm{boundary terms} +
    \int \delta g^{\alpha \beta} \Big[ \partial_{\mu} \Big( \sqrt{-g} g_{\alpha \beta} g^{\mu \nu} \partial_{\nu} f_{, \ourB} \Big) - \partial_{\alpha} \big( \sqrt{-g} \partial_{\beta} f_{, \ourB} \big) \Big] d^4 x \,,
\end{multline}
with all boundary terms proportional to $\delta g_{\mu \nu}$ and $\partial \delta g_{\mu \nu}$ vanishing. Expanding the partial derivatives of~(\ref{integration_by_parts_B_Appendix}) gives
\begin{multline}
  \label{integration_by_parts_final_Appendix}
  \int \delta g^{\alpha \beta} \Big[
    \partial_{\mu} \Big( \sqrt{-g} g_{\alpha \beta} g^{\mu \nu} \partial_{\nu} f_{, \ourB} \Big) -
    \partial_{\alpha} \big( \sqrt{-g} \partial_{\beta} f_{, \ourB} \big)
    \Big]\, d^4 x  \\ =
  \int \sqrt{-g} \delta g^{\rho \sigma} \bigg[
    \partial_{\nu}(f_{, \ourB}) \Big[2 g_{\eta \rho} g^{\mu \nu} \Gamma^{\eta}_{\mu \sigma} -
      g_{\rho \sigma} g^{\mu \eta} \Gamma^{\nu}_{\mu \eta}  \Big] -
    \partial_{\sigma}(f_{,\ourB}) \Gamma^{\eta}_{\rho \eta} \\ +
    \partial_{\mu}\partial_{\nu}(f_{,\ourB})g_{\rho \sigma} g^{\mu \nu} -
    \partial_{\rho} \partial_{\sigma}(f_{,\ourB}) \bigg] \, d^4 x \,.
\end{multline}
Putting equations~(\ref{Integration_by_parts_B_first_term_Appendix}) and~(\ref{integration_by_parts_final_Appendix}) into our expression for the boundary variation~(\ref{variation_of_B_Int_by_Parts}), and cancelling off terms, we arrive at
\begin{multline} 
    \int \sqrt{-g} f_{, \ourB} \delta \ourB  \, d^4 x= \int \delta g^{\rho \sigma} \sqrt{-g} \bigg[ \frac{1}{2} f_{,\ourB}  g_{\rho \sigma} \ourB 
     - \partial_{\nu}(f_{,\ourB}) \Big[ \frac{1}{2} g_{\rho \sigma} g^{\mu \nu} \Gamma^{\eta}_{\mu \eta} + \frac{1}{2} g_{\rho \sigma} g^{\mu \eta} \Gamma^{\nu}_{\mu \eta} - \delta^{\nu}_{\sigma} \Gamma^{\gamma}_{\rho \gamma} \Big]  \\
    + \partial_{\mu}\partial_{\nu}(f_{,\ourB})g_{\rho \sigma} g^{\mu \nu} - \partial_{\rho} \partial_{\sigma}(f_{,\ourB})
     \bigg]\, d^4 x \,.
    \end{multline}
Lastly, let us rewrite the connection pieces in terms of partial derivatives of the metric
\begin{multline}
  \label{boundary_term_variation_final_appendix}
  \int \sqrt{-g} f_{, \ourB} \delta \ourB  \, d^4 x =
  \int \sqrt{-g}  \delta g^{\rho \sigma} \bigg[
    \frac{1}{2} f_{, \ourB}  \ g_{\rho \sigma} \ourB  + g_{\rho \sigma}\partial^{\mu}
    \partial_{\mu}  f_{, \ourB} - \partial_{\rho} \partial_{\sigma} f_{, \ourB} \\ +
    \frac{1}{2} g_{\rho \sigma} \partial_{\mu}(g^{\mu \nu}) \partial_{\nu} f_{, \ourB} +
    \frac{1}{\sqrt{-g}} \partial_{\rho}(\sqrt{-g})\partial_{\sigma} f_{, \ourB} \bigg]\, d^4x \,.
\end{multline}

\subsection{\texorpdfstring{$f(\ourG,\ourB)$}{f(G,B)} field equations} 

The full variation of the gravitational $f(\ourG,\ourB)$ action is given by the variation of the bulk term~(\ref{bulk_term_variation}), the boundary term~(\ref{boundary_term_variation_final_appendix}) and the metric determinant
\begin{multline}
  \delta S_{\rm grav} = \frac{1}{2\kappa} \int \Bigg[ \sqrt{-g} \bigg(  \frac{\partial  f(\ourG,\ourB)}{\partial \ourG} \delta \ourG + \frac{\partial  f(\ourG,\ourB)}{\partial \ourB} \delta \ourB \bigg)  - \frac{1}{2}  f(\ourG,\ourB)\sqrt{-g} g_{\rho \sigma} \delta g^{\rho \sigma} \Bigg]\, d^4 x \\
  = \frac{1}{2\kappa} \int \sqrt{-g} \delta g^{\rho \sigma} \Bigg[ f_{, \ourG} \Big( G_{\rho \sigma} + \frac{1}{2} g_{\rho \sigma} \ourG \big) + \frac{1}{2} E_{\rho \sigma}{}^{\gamma} \partial_{\gamma} f_{, \ourG} - \frac{1}{2}   g_{\rho \sigma} f(\ourG,\ourB)  \\
    + \frac{1}{2}g_{\rho \sigma}  f_{, \ourB}  \ourB  + g_{\rho \sigma}\partial^{\mu}  \partial_{\mu}  f_{, \ourB}
    - \partial_{\rho} \partial_{\sigma} f_{, \ourB}
    + \frac{1}{2} g_{\rho \sigma} \partial_{\mu}(g^{\mu \nu}) \partial_{\nu} f_{, \ourB}+ \frac{1}{\sqrt{-g}} \partial_{\rho}(\sqrt{-g})\partial_{\sigma} \ f_{, \ourB}   \Bigg] \, d^4 x \,.
\end{multline}
Including the matter action $S_{\rm{matter}}[g_{\mu \nu},\Phi]$, the variation of the total action $S_{\rm{total}} = S_{\rm grav} + S_{\rm{matter}}$ leads to the field equations
\begin{multline}
  \label{Final_field_equation_Appendix}
  \frac{\partial f}{\partial \ourG}\Big( G_{\rho \sigma} + \frac{1}{2} g_{\rho \sigma} \ourG \big) + \frac{1}{2} E_{\rho \sigma}{}^{\gamma} \partial_{\gamma} \Big( \frac{\partial f}{\partial \ourG} \Big) - \frac{1}{2}  f(\ourG,\ourB) g_{\rho \sigma} + \frac{1}{2} g_{\rho \sigma} \ourB  \frac{\partial f}{\partial \ourB} \\
  + g_{\rho \sigma}\partial^{\mu}  \partial_{\mu}\Big( \frac{\partial f}{\partial \ourB}\Big)
  - \partial_{\rho} \partial_{\sigma} \Big( \frac{\partial f}{\partial \ourB}\Big)
  + \frac{1}{2} g_{\rho \sigma} \partial_{\mu}(g^{\mu \nu}) \partial_{\nu}\Big(\frac{\partial f}{\partial \ourB}\Big) + \frac{1}{\sqrt{-g}} \partial_{(\rho}(\sqrt{-g})\partial_{\sigma)} \Big(\frac{\partial f}{\partial \ourB} \Big) = \kappa T_{\rho \sigma} \,,
\end{multline}
where the metric energy-momentum tensor is defined as 
\begin{align}
  T_{\mu \nu} = -\frac{2}{\sqrt{-g}}\frac{\delta L_{\rm{M}}[g_{\mu \nu}, \Phi]}{\delta g^{\mu \nu}} \,.
\end{align}
Note that we have explicitly symmetrised over $\rho$ and $\sigma$ in the field equations~(\ref{Final_field_equation_Appendix}).

\addcontentsline{toc}{section}{References}
%\bibliographystyle{jhepmodstyle}
%\bibliography{references}

\providecommand{\href}[2]{#2}\begingroup\raggedright\endgroup

\end{document}